\documentclass[useAMS,usenatbib]{aastex}
\usepackage{graphicx,subfigure}
\usepackage{rotating}
\usepackage{epsfig}
\usepackage{amssymb}
\usepackage{amsmath}
\setcitestyle{numbers}
\usepackage[varg]{txfonts}
\usepackage{graphicx}
\shorttitle{fast transient QPOs in H1743-322}
\shortauthors{Sriram et al.}

\begin{document}
\title{SPECTRO-TEMPORAL STUDIES OF RAPID TRANSITION OF THE QUASI-PERIODIC OSCILLATIONS IN THE BLACK HOLE SOURCE H1743-322}

\author{K. Sriram\altaffilmark{1,2}, S. Harikrishna\altaffilmark{1}, and C. S. Choi\altaffilmark{2} }
\altaffiltext{1}{Department of Astronomy, Osmania University, Hyderabad 500007, India.}
\email{astrosriram@yahoo.co.in}
\altaffiltext{2}{ Korea Astronomy and Space Science Institute, Daejeon 34055, Republic of Korea.}

\begin{abstract}

An appearance or disappearance of QPOs associated with the variation of X-ray flux can be used to
decipher the accretion ejection mechanism of black hole X-ray sources. We searched and studied such rapid transitions in H1743-322 using RXTE archival data and found eight such events, where QPO vanishes suddenly along with the variation of X-ray flux.
The appearance/disappearance of QPOs were associated to the four events exhibiting type-B QPOs at $\sim$ 4.5 Hz, one with type-A QPO at $\nu$ $\sim$ 3.5 Hz, and the remaining three were connected to type-C QPOs at $\sim$ 9.5 Hz. Spectral studies of the data unveiled that an inner disk radius remained at the same location around 2--9 r$_{g}$, depending on the used model but power-law indices were varying, indicating that either corona or
jet is responsible for the events. The probable ejection radii of corona were estimated to be around 4.2--15.4 r$_{g}$ based on the plasma ejection model. Our X-ray and quasi-simultaneous radio correlation studies
suggest that the type-B QPOs are probably related to the precession of a weak jet though a small and weak corona is present at its base
and the type-C QPOs are associated to the base of a relatively strong jet which is acting like a corona. 

\end{abstract}

\section{Introduction}
 A general understanding of the radiative and geometrical structure of the accretion disk in black hole X-ray binaries (BHXBs) exists now, but
finer details such as the Comptonization region/corona, jet, disk winds and their coupled dynamics with the Keplerian portion of
disk is not yet clearly understood. The {\it Rossi X-ray Timing Explorer (RXTE)} results have provided substantial
information on the spectral and temporal evolution of the outbursts exhibited by the BHXBs
(McClintock \& Remillard 2004; Remillard \& McClintock 2006; Done et al. 2007). During the X-ray outburst,
temporal and spectral parameters smoothly traverse across the 'q' shaped Hardness Intensity Diagram (HID) (Belloni et al. 2011).
The quasi simultaneous X-ray--radio observations have enriched the understanding of the inter-relationship between the disk-steady jet and the transient jet
phenomena across the HID (Fender et al. 2004, 2009). The outburst starts with Low hard/Hard state, i.e. right-aligned portion of the HID, where the disk is often considered to be truncated
 with an optically thin, high temperature, Comptonized-zone/corona/sub-Keplerian flow in the inner region of the accretion disk along with a low Lorentz factor ($\Gamma_{jet}$ $<$ 2) steady jet.
 As the X-ray luminosity increases or near the peak of the outburst, the source occupies a large plane in the HID known as the hard intermediate and soft intermediate (Belloni 2010)/steep power-law states (HIMS/SIMS/SPL), where the disk is
 close to the last stable orbit or moderately truncated along with an optically thick, low temperature corona in the inner region.
 The jet line in the HID falls in this state where its activity turns on/off, however, the jets are transient/ballistic in nature with $\Gamma_{jet}$ $>$ 2 (Fender et al. 2004).
 Finally the source occupies the left-aligned portion of the HID with thermal dominated (TD) state properties. In this state, the disk is not truncated, has extremely low Comptonizing emission and the jet ceases to exist.

The X-ray power density spectrum (PDS) of the outburst also follows the HID q-shaped track.
The strong broad band limited noise along with a low frequency quasi periodic oscillation (LFQPO) ($\le$0.1 Hz) is seen at the onset of the outburst and
as the disk approaches towards the BH, the QPO frequency increases until $\sim$10 Hz along with  high-frequency QPOs in HIMS/SIMS.
The PDS becomes weak and QPO is absent in the TD state (Belloni \& Motta 2016).
Based on the frequency and rms, these QPOs are classified into three different types viz. A, B, and C (Wijnands et al. 1999;
Homan et al. 2001; Remillard et al. 2002; Casella et al. 2005). Among these, the appearance of type-B QPO is always associated with
the appearance of a jet line in the HID and on a few occasions, sudden flux variations are also noticed.
Type-B QPO varies in a limited frequency range $\sim$4-6 Hz with a low quality factor (Q={$\frac{\nu}{\delta\nu}$}) often observed at
the peak of the X-ray outburst along with a moderate radio flux. Sudden flux variations or flip-flop transitions associated with type-B/C to type-C/B
are rare and seen in very few BHXBs. These flux variations and type-B/A QPO transitions were seen in GX 339-4 (Miyamoto et al. 1991; Nespoli et al. 2003;
Motta et al. 2011), XTE J1550-564 (Homan et al. 2001; Sriram et al. 2016), XTE J1859+226 (Casella et al. 2004; Sriram et al. 2013),
H1743-322 (Homan et al. 2005), XTE J1817-330 (Sriram et al. 2012), MAXI J1659-152 (Kalamkar et al. 2011; Kuulkers et al. 2013), MAXI J1535-571 (Stevens et al. 2018) with no sudden flux variation and type-C/B QPO transition in MAXI J1820+070 (Homan et al. 2020), and GRS 1915+105 (Soleri et al. 2008). A 45\% sudden flux variation, with an appearance of type-C QPO in the low flux state
that disappeared in the high flux state in 40 s, was observed in the BH source Swift J1658.2-4242 using the data from {\it NuSTAR} and {\it XMM-Newton} (Xu et al. 2019).
 {\it Astrosat} also observed a similar phenomenon at the peak of the outburst and spectral results indicated a slight increase in the inner disk temperature along with no variation in the power-law index for the low and high flux state (Jithesh et al. 2019). Bogensberger et al. (2020) studied all the flip-flop transitions of this source and concluded that change in the inner disk temperature is responsible for the sudden transitions.

The origin of LFQPOs is known to associate with the inner region of the accretion disk, but its precise location is still unknown
and various models predict different occurrence mechanisms. Stella \& Vietri (1998) connected those to the Lense-Thirring (LT) precession at a single radius in the Keplerian portion of
the accretion disk. LT precession of a rigid inner flow can explain the mechanism of LFQPOs (0.1--10 Hz) (Fragile et al. 2007; Ingram et al. 2009; Ingram \& van der Klis 2013). 
 A non-rigid precession was proposed where a differential precession between the disk and jet components is required to explain the energy dependence of QPO (van den Eijnden et al. 2016, 2017).
 The accretion ejection instability (AEI) model explains type-A, B, C QPOs as arising due to the instability which transports energy from the magnetized accretion disk to the inner corona, wherein
 relativistic formalisms are needed to explain type-B and A QPOs (Tagger \& Pellat 1999; Varniere \& Tagger 2002; Varniere \& Vincent 2016).
 Type-B QPOs are often associated with relativistic jets/ejecta (Motta et al. 2015; Fender et al. 2004, 2009), however, it is unclear how the precession of jet or coupled jet-corona is associated with the
 production of type-B QPOs.

Since the sudden appearance/disappearance of QPOs often occurs close to the BH and their connection to the corona or jet is not properly understood,
we study such rapid transition events in the BH source H1743-322. This source was discovered in August 1977 by {\it Ariel} during a bright outburst and
accurately pin-pointed by High Energy Astronomy Observatory 1 (HEAO 1; Kaluzienski \& Holt 1977; Doxsey et al. 1977). It was detected in 1984 by
EXOSAT (Reynolds et al. 1999) and by the TTM/COMIS telescope onboard the Mir-Kvant observatory in 1996 (Emelyanov et al. 2000). But after its rediscovery by INTEGRAL on 2003 March 21 (Revnivtsev et al. 2003),
extensive observations were made a few days later both in X-ray by {\it RXTE} (Markwardt \& Swank 2003) and in radio by Very Large Array (VLA; Rupen et al. 2003).
Relativistic jets were observed as plasma blob ejections during the 2003 outburst, thus making it a micro-quasar (Corbel et al. 2005).
After 2003, the source exhibited smaller outbursts from 2004 to 2011 (Kuulkers et al. 2008; Krimm et al. 2009; Zhou et al. 2013). This source is located at a
distance of $\sim$ 8.5 kpc with an inclination {\it i}$\sim$75$^{\circ}$ and a mass of the BH is estimated to be 5--15 M$_{\odot}$ with a spin of a$_*$ = -0.25--0.75 (Steiner et al. 2012).
It was also detected in optical and infrared wavelengths (Baba et al. 2003; Steeghs et al. 2003; Chaty et al. 2015) and the donor was found to be a late type star located in the galactic bulge.
Apart from the detection of characteristic LFQPOs, a pair of high frequency QPOs were reported at $\sim$ 160 Hz and $\sim$ 240 Hz by Homan et al. (2005) and $\sim$ 166 Hz and $\sim$ 242 Hz by Remillard et al. (2006).
McClintock et al. (2009) extensively studied the 2003 outburst in X-ray, radio, \& optical band and concluded that its properties closely match with the XTE J1550-564.

\section{Data Reduction and Analysis}

We used the {\it RXTE} Proportional Counter Array (PCA,  Jahoda et al. 2006) archival data for
H1734-322 to search for sudden transition events during the outbursts in the years 2003, 2004, 2005, 2007/2008, 2008, 2009, 2009/2010 and 2011 (Markwardt \& Swank 2003; Belloni et al. 2008; Krimm et al. 2009; Zhou et al. 2013) and after that RXTE was decommissioned on 2012 January 5. The longest outburst being the one in 2003 where 170 pointed
observations were made with RXTE (see McClintock et al. 2009). Previous studies indicate that the sudden transition occurred around peak of the outbursts in BHXBs and most of them are associated with transition or disappearance of 3--8 Hz QPOs  (eg. Sriram et al. 2016). Hence we have selected 103 observations around the peak of the outbursts associated with the QPOs exhibited by the source. We extracted the PDS for all the observations and looked for the appearance/disappearance of the QPOs in these observations and marked the transitions in the light curves by finely studying the PDS with a time resolution of 20 s. So we have selected the PCA data viz. standard 2 and single bit modes were used to obtain the light curves and spectra.
Power density spectra (PDS) were extracted (time bin = $\frac{1}{512}$ s) during sudden flux variations. On eight occasions, QPOs suddenly appeared/disappeared as the
flux varied. One such event, where a type-C QPO rapidly varied to type-B QPO, was reported by Homan et al. (2005) but spectral analysis was not reported.
In order to fit the PDS in units of (rms/mean)$^{2}$ Hz$^{-1}$ with white noise level subtracted, a model consisting of {\it power-law + two/one Lorentzian} was invoked
to evaluate the QPO parameters.
For these events, the best-calibrated PCU2 unit spectra were extracted in the 3.0--25.0 keV energy band.
The updated background model, PCA history file spanning the entire mission, appropriate response matrix along with the calibrated database were used
for extracting the spectra \footnote{heasarc.gsfc.nasa.gov/docs/xte/pcanews.html}.
A systematic error of 0.5\% was added to the respective spectra. HEASOFT v6.19 software sub-packages were utilized to reduce the data. The XSPEC v12.9 (Arnaud 1996) was used for spectral analysis and parameter uncertainties were calculated at the 90\% confidence level i.e. $\Delta$ $\chi^2$ = 2.71 or otherwise mentioned.

\section{Timing Analysis}
 A search for events associated with sudden transitions of QPOs in H1743-322 was undertaken. 
A total of eight observations were found (see Table 1) which exhibited sudden variations in PDS and flux. The occurrence of type-A \& type-B (filled circle) and type-C (open circle) QPOs are marked in the ASM light curves of 2003 and 2009 outburst (Figure 1) and
most of them occurred at the peak of the outbursts. We did not find such sudden variations in other outbursts exhibited by H1743--322. McClintock et al. (2009) classified these observations in SPL state and more specifically, type-B and type-A QPOs are associated with SIMS and type-C QPOs belong to HIMS. Figure 2 and 3 show the light curves in 6.12--14.76 keV energy band, vertical lines
show the transitions where QPO suddenly appeared/disappeared represented with Q (QPO) and NQ (non-QPO) sections. Here the appearance means that QPO is present in the PDS of the respective section of the light curve and disappearance means that the QPO is not present.  Power density spectra (PDS) were extracted in 3.68--5.71 keV and 6.12--14.76 keV bands and the results of the fits are presented in Table 2. Out of the eight, four events were found to be displaying QPOs at $\nu$ $\sim$ 4.5 Hz along with a harmonic, one event with QPO at $\nu$ $\sim$ 3.5 Hz, and the remaining three displayed QPO at $\nu$ $\sim$ 9.5 Hz with no harmonic (see Table 2).
QPOs with $\nu$ $\sim$4.5 Hz have properties of type-B QPO (Casella et al. 2005; Motta et al. 2015) with a quality factor Q = {$\frac{\nu}{\delta \nu}$ $<$ 5 and fractional rms amplitude $<$ 5\% in the energy band 6.12--14.76 keV except in ObsID 80135-02-02-000, whereas QPO at $\nu$ $\sim$ 3.5 Hz can be classified as type-A QPO (Q = 2.48) and QPOs at $\nu$ $\sim$ 9.5 Hz have Q $\sim$ 8 with rms 4.7--6.1\%.
Though Motta et al. (2015) have not reported QPOs in ObsID 80146-01-17-00 and 80146-01-25-00 but found type-B QPOs in ObsID 80146-01-16-00 and 80146-01-26-00. The later observations QPO parameters were found to be similar in nature with the QPO parameters of earlier observations as shown in Table 2.   

 In ObsID 80135-02-02-000, a type-C QPO at $\nu$ $\sim$ 4.65 Hz
 with Q = 4.8 along with a harmonic at $\nu$ = 9.28 Hz suddenly varied to a type-B QPO with no harmonic at $\nu$ $\sim$ 5.57 with a Q = 3.5 and this
 feature remained unchanged for the other three sections (not shown here; see Homan et al. 2005). The type-B QPOs are detected at a significance level of 6--8 $\sigma$ and 2.8 $\sigma$ for type-A QPO. In the three other observations,
 a type-C QPO around $\sim$ 9.5 Hz disappeared in the PDS as the source varied rapidly. The quality factor of type-C QPO is relatively higher than that of type-B QPO, a primary characteristic of type-C QPOs and they are detected at a significance level of 8-10 $\sigma$. Moreover for a few observations, 
 QPOs were absent in the low energy band but present in the higher energy bands. For one such observation the absence of QPO is shown in Figure 3 for ObsID 80146-01-13-00.

\section{Spectral Analysis}

We performed spectral analysis for all the observations where QPOs either appeared or disappeared, and on one occasion, transitioned from type-C to type-B.
We used three different types of models to unfold the spectra in the energy band 3.0--25.0 keV, viz. model 1: {\it wabs$\times$(diskbb+power-law}) (diskbb; Makishima et al. 1986),
model 2: {\it wabs$\times$(diskbb+broken power-law}) based on McClintock et al. (2009) and model 3: {\it wabs$\times$(diskbb+CompTT}).
For the fits, the hydrogen equivalent column density was fixed at N$_{H}$ = 2.2 $\times$ 10$^{22}$ cm$^{-2}$ (McClintock et al. 2009). A few observations exhibited Fe line which was modeled using a Gaussian function with a
centroid energy E$_{c}$ = 6.4 keV (fixed).
The CompTT model has four parameters viz. input soft seed photon temperature kT$_{s}$, electron temperature kT$_{e}$, optical depth $\tau$, and normalization (Titarchuk 1994). A spherical geometry of the plasma cloud was
adopted for the fits. During the fits of model 3, the kT$_{s}$ parameter was allowed to vary but in order to constrain the error bars, it was later fixed at its best-fit value otherwise this parameter is pegged at zero during the fits.

It was found that observations with $\nu$ $\sim$ 4.5 Hz \& 3.5 Hz QPOs fitted well when unfolded with model 1 (Fig. 4). Disk normalization (N$_{diskbb}$) was found to be varying in all these observations along with the power-law index ($\Gamma$) from Q to NQ sections (Table 3).
For example, in ObsID 80146-01-17-00, the change in N$_{diskbb}$ is required at a F-test probability 6.9$\times$ 10$^{-30}$.
All the observations displayed reasonably good fits when unfolded with model 2 (Table 3 and Table 4). The broken power-law index ($\Gamma_{1}$) was observed to be varying from Q to NQ
sections in all the observations, clearly indicating towards variation in plasma properties in the inner region of the accretion disk. $\Gamma_{2}$ was found to be changing in two observations only and it
remained consistent in other observations (see Table 3 \& 4). In the observations with QPOs around $\nu$ $\sim$9.5 Hz, model 3 resulted in a better fit than those of model 1 and 2 (e.g., for ObsID 80146-01-50-00, model 1 fit gave $\chi^{2}$/dof = 70/44 and model 2 $\chi^{2}$/dof = 26.5/40).
The electron temperature kT$_{e}$ was found to be varying in two of the observations ($\ge$ 2 keV) along with N$_{diskbb}$, however it is to be noted that these changes are not explicit as they have obtained after freezing the kT$_{s}$ parameter (Fig. 5, Table 4 ).
It was noted that the spectra unfolded with model 3 were resulting in unrealistically high electron temperature for the observations associated with QPOs of $\sim$4.5 Hz
(e.g., for ObsID 80146-01-17-00 kT$_{e}$ $\sim$ 300 keV). Even when kT$_{e}$ was fixed at 10 keV, error bars for the other spectral parameters could not be constrained.

In ObsID 80135-02-02-000 (a type-C QPO transitioned to type-B QPO), based on model 1, the spectral fit suggests that the inner disk radius has slightly decreased but $\Gamma$ remained the same at 2.65 (Table 3).
Whereas when unfolded with model 2, the broken power-law index was found to be decreasing ($\Gamma_{1}$ = 2.48 $\pm$ 0.01 to 2.33 $\pm$ 0.02) along with no consistent change in N$_{diskbb}$ and flux was found to be increasing during the transition.

\section{Robustness of Observed Spectral Variation}

In order to check the significance of the spectral variations in Q and NQ and in A and B sections (Figure 2 and 3), we fitted these sections spectra simultaneously with the models discussed above and also with wabs$\times$(SIMPL*diskbb).  It was observed that three observations viz. 80146-01-17-00, 80146-01-25-00, and 80146-01-51-01 exhibit changes in the power-law index of {\it SIMPL} model between Q and NQ. The parameter f$_{scat}$ was found to be varying for example in ObsID 80135-02-02-000 and the variation of the disk normalization parameter was noticeable in ObsID 80146-01-13-00. In order to see the significance of the variation among these parameters, STEPPAR function in XSPEC was invoked and confidence contours with 68\%, 90\% and 99\% (Figure 6) were obtained for different observation sections. It can be seen that all the changes are significant at the 99\% confidence level.

Moreover, we performed F-test probability in order to check which spectral parameter is primarily varying from one section to another. For this procedure, we tied the spectral parameters of Q and NQ sections and freeze the parameters as the best-fit values of NQ section spectrum. Then we allowed to vary the parameters independently one by one as shown in Table 6. Figure 7 (left) shows the square root of $\chi^{2}$ variation as the spectral parameters were independently varying in a stepwise manner. It can be seen that the residuals of other sections gradually approaching towards their best-fit values as demanded by the data. The low value of F-test probability for ObsID 80146-01-17-00 was observed at $\Gamma_1$ parameter (2.15 $\times$ 10$^{-94}$) which indicates that it is the primary parameter which needs to be varying in order to explain the spectral difference between NQ and Q sections, similarly for ObsID 80135-02-02-000 $f_{scat}$ is the primary parameter ( 6.54 $\times$ 10$^{-51}$ ).

\section{Discussion and Results}
The study of rapid appearance/disappearance of QPOs is important in order to understand the disk-jet ejection or quenching phenomena in the inner accretion disk of BHXBs.
In H1743-322, we found total eight events where the QPO has disappeared/appeared in different observations. The PDS study clearly suggests that they were type-B QPOs with $\nu$ $\sim$ 4.5 Hz, one type-A $\nu$ $\sim$ QPO 3.5 Hz, and type-C QPO with $\nu$ $\sim$ 9.5 Hz (Table 2).
 A rapid emergence of type-B QPO around $\sim$ 5-7 Hz after a type-A QPO within 10 s was noticed in GX 339-4 and was related to the spectral hardening of the flux  (Nespoli et al. 2003). The power-law index was found to vary from 2.46$\pm$0.03 to 2.29$\pm$0.03 as a type-B QPO at $\sim$ 5.54 Hz suddenly transitioned to a type-A QPO in a few tens of seconds in XTE J1817-330 (Sriram et al. 2012). In case of XTE J1859+226 where type-B QPOs $\sim$ 6 Hz varied to type-A \& C QPOs ($\sim$ 7.6 -- 8.7 Hz) in different observations, the soft and hard fluxes varied by 3--20 \% (Sriram et al. 2013). In one of the observations, a type-A QPO $\sim$ 7.6 Hz suddenly appeared along with the hardening of the spectra ($\Gamma$ $\sim$ 2.40 to $\Gamma$ $\sim$ 2.27 ) (Sriram et al. 2013). A type-A QPO varied to type-B QPO at $\sim$ 3 Hz with an harmonic $\sim$ 6 Hz  was noticed in XTE J1550-564 and both soft and hard components were found to change during the transition (Sriram et al. 2016). An appearance of type-A(or type-B) QPO at 5.72 Hz was seen in MAXI J1535--571 which occurred in a duration of a few days (Stevens et al. 2018). In case of Swift J1658.2-4242, a decrease in the X-ray flux variation by 45\% was noted with a sudden production of type-C QPO in 40 s (Xu et al. 2019). In the same source, type-C to type-A QPO were found to be related to rapid flip-flop transitions which were found to be associated with the change of inner disk temperature (Bogensberger et al. 2020). A type-C QPO along with an harmonic $\sim$ 4.5 Hz and 9 Hz varied to type-B QPO at $\sim$ 3 Hz in a duration of a few thousand seconds was observed in MAXI J1820+070, however no spectral studies were reported (Homan et al. 2020).


We found changes in kT$_{in}$ and N$_{diskbb}$ in a few observations associated with the type-B and type-C QPOs. Since the true radius of the inner disk (R$_{in}$) is affected by
scattering and spectral hardening (Merloni et al. 2000; Done \& Davis 2008), we used the relation 1.2$\times$(N$_{diskbb}$/cos $\theta$)$^{1/2}$ $\times$ (D/10 kpc) km to derive the true inner disk radius R$_{in}$ (Reynolds \& Miller 2013).
An inclination of 75$^\circ$ and a distance of 8.5 kpc was used to obtain R$_{in}$ in different observations for Q and NQ sections (Steiner et al. 2012; see Table 3 and Table 4).
We noted that during the occurrence of type-B and type-C QPOs , R$_{in}$ is similar to each other with an uncertainty of $\sim$ 10 km ($\sim$ 1 r$_{g}$, gravitational radius r$_{g}$ = G M / c$^{2}$). In the truncated accretion disk scenario
(Done et al. 2007), the lower frequency QPO should be arising from an outer region when compared to the high frequency QPO. But here it should be noted that the origin of type-B QPOs are
different from type-C QPOs (Motta et al. 2011; Sriram et al. 2013; Belloni \& Motta 2016). Such a similar radius (2--3 r$_{g}$ $\sim$ 23--34 km assuming a 7 M$_{\odot}$ BH for a spin parameter a$_{*}$ = 0.2)
is predicted by the Jet Emitting Disk (JED) model in the inner region of the accretion disk (Marcel et al. 2020).

Using model 1, $\Gamma$ was found to be softer whenever a type-B QPO was observed (e.g. ObsID 80146-01-25-00 for Q, $\Gamma$ = 2.83$\pm$0.03 and for NQ $\Gamma$ = 2.69$\pm$0.03 see Table 3).
The change in power-law flux is relatively high when compared to diskbb flux. Similarly, model 2 exhibited steeper (soft) power-law indices $\Gamma_{1}$ for three observations 80146-01-17-00, 80135-02-02-000 \& 80146-01-51-01 (Q \& A sections) and harder indices for others, whereas $\Gamma_{2}$ also varied significantly in three observations (Table 3). 
This can be understood in the framework of a multi-component hot flow where the outer region of the hot Compton component (i.e. soft Compton component) excites the harmonic and the
inner hot flow/jet is responsible for the fundamental QPO based on the frequency resolved spectroscopy (Axelsson et al. 2013; Axelsson \&  Done 2016; Hjalmarsdotter et al. 2016).
We hypothesize that $\Gamma_{1}$ is arising from the outer hot flow and $\Gamma_{2}$ is associated with the inner hot flow, probably the base of jet.
In our study, we also detected disappearing harmonics along with the fundamental QPO indicating that the some portion of these flows are ejected away.
On one occasion (ObsID 80135-02-02-000) a type-C QPO along with a harmonic transitioned to a type-B QPO where a slight change in R$_{in}$ was noted,
however no consistent change was noted in $\Gamma$ (model 1).  In model 2, $\Gamma_{1}$ has significantly changed but no variations were seen in $\Gamma_{2}$ and R$_{in}$. The broken power-law flux increased by 45\% and soft flux varied by 2\% (Table 3). The variation in $\Gamma_{1}$ indicates that the outer region of hot flow was ejected away leaving behind the inner hot flow or a jet.

Most of the observations associated with the type-B QPO strongly indicate toward the precession of jet or a vertically extended corona structure (Ingram et al. 2009; Motta et al. 2011, 2015; Sriram et al. 2016; Xu et al. 2019; Jithesh et al. 2019; Bogensberger et al. 2020; Belloni et al. 2020).
A phase resolved spectroscopy of type-B QPO in GX 339-4 vividly indicated that a precession of a large scale height probably a base of jet is needed to explain the non-phase modulation in the black-body and Comptonized emissions (Stevens \& Uttley 2016).
A detailed study of phase difference in QPO harmonics in BHXBs led to the conclusion that the type-B QPO phase difference remains constant around 0.5--0.6 for most of the sources, however, it varies and increases with frequency for type-C QPOs.
In order to explain this scenario, a precessing optically thick jet is required to produce the type-B QPO and the associated harmonic is due to the non-sinusoidal nature of the fundamental wave (de Ruiter et al. 2019).
Type-B QPO does not show any coupling between the QPO and the broadband noise and it does not exhibit any inclination dependency (Arur \& Maccarone 2019, 2020).
Motta et al. (2015) found that type-B QPOs are stronger in low inclination systems and are associated with a different physical origin than that of type-C QPOs.
They are somehow possibly connected to the jet whose power is stronger in face-on systems. This is also supported by the X-ray and radio correlation studies (Motta et al. 2018).

During the sudden disappearance of type-C QPOs, for example in ObsID 80146-01-13-00, kT$_{in}$ varied from 1.32 keV to 1.47 keV along with a consistent variation in
R$_{in}$, but kT$_{e}$ did not fluctuate much (model 3). However, kT$_{e}$ varied significantly in other two observations when $kT_{s}$ is fixed and similar changes were observed from model 2 (Table 4). In Table 4, for model 3 power-law flux increased by 38 \% and disk flux increased by 9 \% and for model 2, 32 \% and 11 \%, respectively.
In all the three observations, the change in hard flux is relatively higher than the soft flux.
This low temperature, high optical depth corona is possibly causing the type-C QPO, a characteristic of this spectral state (Done et al. 2007; Sriram et al. 2009). The disappearance of QPO could be due to a partial ejection
of this component, leaving behind a relatively hotter component which is evident from the spectral results (model 3, Table 4). In all the three observations, the $\Gamma_{1}$ found to be relatively steeper associated with Q sections (variation was significant in one observation) along with a decrease in kT$_{in}$ and no consistent variation in $\Gamma_{2}$ was observed between Q and NQ sections (Table 4).

One of the most promising models which can explain the type-C QPO is the Lense-Thirring precessing model where the inner hot region precesses like a rigid body. Strong evidence associated with the Lense-Thirring precession model came from the detection of quasi periodic modulation of the iron line in H1743-322 and the line modulation is due to the quasi periodic illumination of the disk caused by the precession of the hot inner flow (Ingram et al. 2009; Ingram \& Done 2012).
 This model clearly suggests that there should exist a truncation radius beyond which the disk does not precesses. The bending wave sets the inner radius at $\sim$ 8 r$_{g}$ and $\sim$ 10 r$_{g}$ depending on the spin of the black hole a$_{*}$ = 0.5 and 0.9 respectively. H/R = 0.2 (ratio between vertical height and radius of the accretion disk) and an outer radius of $\sim$ 6.5--50 r$_{g}$ can successfully produce the LFQPOs ( 0.1--10 Hz) observed in BHXBs ( Lubow et al. 2002; Fragile et al. 2007; Ingram et al. 2009).
As discussed above most of the observational evidence related to the type-B QPO lead to a possible jet origin and a precessing jet in the inner region.
This model also warrants a truncation in the disk, possibly arising due to the base of a jet, with the jet being the source of hard X-rays via comptonization (Liska et al. 2018; Kylafis et al. 2020; Kylafis \& Reig 2018; Reig \& Kylafis 2015, 2019).

\section{Constraining the Inner Radius Using SIMPL Model}

Spectral results indicate that the disk is almost at the last stable orbit 2--4 r$_{g}$ and exactly determining the inner disk
radius is difficult because of various reasons like color correction and distance. Also during the spectral fitting, the power-law index
diverged in the softer portion of the spectra which affects the inner disk radius (Steiner et al. 2009). In order to overcome this problem,
we used a model wabs$\times$(SIMPL*diskbb) to unfold the spectra of Q and NQ sections (Table 5). This model gave a reasonable fit ($\chi^2$/dof) and
relatively higher inner disk radius when compared to the radius derived from the earlier mentioned models (for more details see Steiner et al. 2009).
Soft photons were up-scattered, resulting in f$_{scat}$ = 0.10--0.55 and the power-law index of SIMPL model ($\Gamma_{simpl}$) was found to be
varying from 2.32 to 3.05 (see Table 5). It was noticed that whenever the type-B QPO of $\nu$ $\sim$ 4.5 Hz was present, it is associated with
a softer $\Gamma_{simpl}$ and for a non-QPO section, a harder ($\Gamma_{simpl}$) was noted, as observed in the other models too.
We found no noticeable variation in the inner disk radius during the sudden disappearance of QPOs except ObsID 80146-01-13-00, where R$_{in}$ varied by at least 6 km (see Table 5).
In ObsID 80135-02-02-000 (type-C to type-B), we found no appreciable change in the $\Gamma_{simpl}$ as well as the disk parameters.
The only parameter that varied was the scattering fraction f$_{scat}$, for the soft photons converting into the power-law component, which varied from 0.16 to 0.22, a 38\% change.
Since there is no noticeable change in the R$_{in}$ and $\Gamma_{simpl}$, the increase in f$_{scat}$ along with the flux can be explained if the vertical structure of the jet has increased in height
 proportionally intercepting relatively more number of soft photons from the disk. Similar non-movement of inner disk radius came from the study of MAXI J1659--152 where type-B QPOs varied from 1.6--4.1 Hz but movement of R$_{in}$ was not noticed from spectral study, clearly indicating that these QPOs are non-disk in origin (Yamaoka et al. 2012).
Similar spectral studies also suggest that R$_{in}$ does not vary much during the emergence of type-B QPOs (Nespoli et al. 2003; Sriram et al. 2012, 2013, 2016; Xu et al. 2019; Jithesh et al. 2019; Bogensberger et al. 2020).
Results from model 2 did not exhibit any significant variation in the R$_{in}$ in four of the observations (Table 3).

\section{QPOs Disappearance and Possible Ejection Heights }

As noticed in PDS both type-B and type-C QPOs are disappearing, suggesting that jet is changing its properties or the inner hot flow is being ejected away.
Here we constrain the vertical height of the inner region using the plasma ejection model (Beloborodov 1999) as most of the observations connected to the type-B QPO
point towards a phenomenon of jet ejection during the outburst (Fender et al. 2009; Motta et al. 2011). In the ejection model, the power-law index is related to the out-flowing
ejecta along the jet $\Gamma$ = 1.9 / B$^{1/2}$, B = $\gamma$ (1+$\beta$) ($\beta$ = {\it v/c} and $\gamma$ = 1 / (1 - $\beta$$^2$)$^{0.5}$. The radio observation of H1732-322
displayed different apparent bulk velocities during different outbursts (for more details see McClintock et al. 2009). Based on the radio observations, Corbel et al. (2005)
found $\beta$ = 0.79 for 2003 outburst and $>$ 0.57 based on decelerating jet knots, while Miller-Jones et al. (2012) constrained $\beta$ to be 0.19--0.28 for the 2009 outburst.
Assuming $\beta$ = 0.79 as observed in the 2003 outburst, the ejection height should be around $\sim$ 3 r$_{g}$ and based on $\Gamma$ $\sim$ 2.3--2.9 from spectral fits, the ejection must
occur at a height of 15.4--4.2 r$_{g}$ but the time of radio ejection does not correspond to the time associated with the X-ray events reported in the present work. For $\beta$ = 0.19--0.28, the ejection heights were found to be $\sim$ 50--25 r$_{g}$, 
however beta for radio ejection occurred on MJD 54984.9 which was $\sim$ 5.36 days earlier than the rapid transition of type-A QPO in the present work (see Table 7).} The heights were calculated assuming the velocities v = $\beta$ $\times$ c to be escape velocities. In the case of $\sim$ 9.5 Hz QPO observations, neither disk parameters nor $\Gamma_{simpl}$ varied significantly, however, f$_{scat}$ was found to be increasing
along with the flux indicating that a jet was present. Now we can assume that during the type-C QPO occurrence, there was a precessing hot corona (Ingram et al. 2009) and this corona was extended
in the vertical direction due to an ejection/transient jet intercepting more number of soft photons and relatively emitting high flux. Theoretically, the presence of a strong steady jet during this transition can be
ruled out because of high H/R, sustaining a strong jet is not possible (King \& Nixon 2018).

\section{Transient QPOs, Its Association to Radio Emission and Jet }

Although there was no radio emission on MJD 52756 and 52763, radio emission was noticed before and after these dates and an appearance
and disappearance of type-B QPOs around $\sim$ 4.5 Hz along with a harmonic was observed in X-rays (see Table 7). During MJD 52788.37, again a radio flux
of $\sim$ 5.59 mJy at 8.46 GHz was observed and no radio observation was reported for MJD 52787.24. The type-B QPO  at 4.2 Hz weakened on MJD 52788.50.
There was no VLA observation on MJD 52751, but on MJD 52752.53, a radio emission ($\sim$ 12.20mJy at 4.86 GHz) was detected which is much lower than that on MJD 52750.60 ($\sim$ 23.10 mJy at 4.86 GHz).
The spectral index, $\alpha$ on MJD 52752.53 was -0.11 (McClintock et al. 2009). QPO at 9.66 Hz disappeared on MJD 52752.97. It clearly suggests that there is a non-steady jet associated
with a variable radio emission. X-ray emission was noticed on MJD 52786.35 (with a disappearance of $\sim$ 9.5 Hz QPO) along with a consistent radio emission during the same time (MJD 52786.35)
with $\sim$ 16 mJy at 8.46 GHz (McClintock et al. 2009) and a spectral index , $\alpha$ = -0.64. During MJD 54990.26, the source was in SIMS (Motta et al. 2010;  Radhika et al. 2016) and our analysis
shows that a type-A QPO $\sim$ 3.5 Hz has disappeared and radio flux density is noted to be $<$1.0 mJy based on the VLA observations (Miller-Jones et al. 2012).
There was an ejection event on MJD 54989.2 with an inverted spectrum ($\alpha$ $\sim$ 1.4 $\pm$0.2) and radio brightness on MJD 54991.4 was interpreted as decay from the original flare.
We also noted that whenever type-B QPOs appeared/disappeared, radio emission was found to be $<$ 10 mJy at 8.46 GHz whereas transition connected with a type-C QPO at 9.5 Hz is associated with a radio emission $>$ 11 mJy. This suggests that a weaker radio jet is probably connected to type-B QPOs. In the table A5 of McClintock et al. (2009), spectral index is found to be negative whenever there is a transition of QPO reported in the present work. Espinasse \& Fender (2018) proposed that negative $\alpha$ is often associated with a radio quiet subset of BHXBs and discrete ejecta evolving between thick and thin states of emission  are probably responsible for the observed spectral indices, however, more studies are needed to confirm this scenario. 
Since simultaneous radio and X-ray observations are not available during the transitions of QPOs and hence precise conclusion can't be made related to the variation of radio emission during these events. 
Table 7 shows the minimum time difference between the X-ray and radio emission during these observations.

The presence of a jet is evident from the radio observations for at least a few observations (Table 7). The inner disk radius during these
events was found to be around $\sim$ 2--6 r$_{g}$ (Table 3 \& 4, model 2), indicating that the disk is close to the last stable orbit. Type-C QPOs most probably are
caused by the LT precession with a rigid inner flow or with a tilted thick disk--jet like structure (Ingram et al. 2009; Liska et al. 2018).
The base of the jet is probably coupled to the precessing inner flow, causing the jet to precess. We found that during the sudden
appearance/disappearance of the observed QPOs, the jet was present, though it is difficult to say which physical component i.e. a rigid inner flow or a jet is connected to these QPOs and in what proportion.
To produce the type-C QPO at 4.5 Hz, in LT precession geometry, the rigid inner flow requires an inner radius of r$_{i}$ = 6.7 r$_{g}$ along with an outer radius of r$_{o}$ = 10.5 r$_{g}$ (for $\zeta$ = -0.5, $\Sigma$ (surface density) $\propto$ r$^{- \zeta}$, spin a = 0.3, H/R = 0.2),
estimated from the formula reported by Ingram et al. (2009). Similarly for the QPO at 9.5 Hz, r$_{i}$= 5.5 r$_{g}$ and r$_{o}$ = 6.1 r$_{g}$ for a=0.2. This theory successfully explains the
production mechanism of type-C QPOs but not type-B QPOs which are often connected to a jet-like structure (Fender et al. 2009; Motta et al. 2011). One of the main differences between type-C and type-B lies
in the quality factor (Q), which is lower in the later case, indicating that a precession of a non-rigid structure is possibly causing the low Q. Such a structure could be a jet which can not be precessing like a rigid body over large scale heights.
Kylafis et al. (2020) showed that a type-B QPO is solely due to the precession of a jet, although a small precessing rigid inner flow is present, and type-C QPO is due to the precessing rigid inner flow despite the presence of a jet.
From our study, one possible evidence for the inner rigid flow is the presence of a low temperature high optical depth compact corona as observed from the spectral fits where type-C QPOs are present (Table 4; Done et al. 2007). A
similar physical component was not observed during the presence of type-B QPOs (for more details see Section 4).

Moreover in their study, $\Gamma$ was found to vary around 2.6--3.0 for an observation angle {\it i}$\sim$75$^{\circ}$ and radius of jet R$_{o}$ = 150--60 r$_{g}$ (see Fig. 2 of Kylafis et al. 2020).
The type-B QPOs presented in our study mostly have $\Gamma$ around $\sim$ 2.5--2.9 (see Table 3), which mandates R$_{o}$ $\le$ 6 r$_{g}$, and are close to the reported inner disk radius $\sim$ 3--6 r$_{g}$. Recently Marcel et al. (2020) have found that type-B QPOs require a  small transition radius (r$_{J}$, truncation radius due to a jet in the inner region) around 2.5 r$_{g}$ but a relativistic treatment is needed in order to fully understand type-B QPOs in their Jet Emitting Disk (JED) model (Ferreira et al. 2006).
At this radius ($<$ 10 r$_{g}$), a precessing ring-like structure which is broken and disassociated with the adjacent disk is also predicted in GRMHD simulations (Nixon et al. 2012, 2013; Nealon et al. 2015; Liska et al. 2020).
No disk breaking is observed in MHD and HD models, but more studies are required in this direction (Hawley \& Krolik 2019). A precessing jet by a tilted black hole disk is also possible at
a radius $\le$ 10 r$_{g}$ (Liska et al. 2018) and if a corona is considered along with a disk and jet configuration, corona flux lags the disk's (Liska et al. 2019).
In their weak magnetic field approximation GRMHD simulation, a jet extends to $<$ 100 r$_{g}$ in one direction and stalls as it torques with ambient medium for a given set of initial parameters (Liska et al. 2019).
King \& Nixon (2018) found that strong jets are not possible for thick disks (H/R $\sim$ 0.1--0.3) but simulations have shown that jets are possible with H/R = 0.3, however, the strength of the jet depends on the magnetic field (Liska et al. 2018).

\section{Conclusion}
1. Eight observations displayed a sudden appearance or disappearance of QPOs, of which four were associated with type-B QPOs at $\sim$ 4.5 Hz, one with type-A QPO at $\nu$ $\sim$ 3.5 Hz, and remaining were type-C QPOs exhibiting $\sim$ 9.5 Hz.

2. Whenever a type-B QPO at $\sim$ 4.5 Hz disappeared, the power-law index varied, vividly indicating that a transient jet is responsible for the accretion-ejection phenomenon.
The presence of a jet is supported by the detection of quasi-simulations radio emission as discussed above. The sudden disappearance of QPOs indicates that some portion of the corona might have ejected away at a height
of 4.2--15.4 r$_{g}$ in the form of a transient jet.

3. The ObsID 80135-02-02-000, where a type-C QPO along with a harmonic varied to a type-B QPO along with the disappearance of the harmonic, is associated with an increase in the scattered soft photon fraction (f$_{scat}$) and flux.
This is possible if the inner corona or a jet-like structure has extended in height, which can intercept more number of soft photons.

4. During the observations, the inner disk radius was around $\sim$ 2--6 r$_{g}$ (Table 3 \&4) and $\sim$ 4--9 r$_{g}$ based on the spectral fits using SIMPL model, we did not notice any sudden variation in R$_{in}$ during the transitions except in one observation (Table 5) which was associated to the type-C QPO.

5. During the disappearance of type-C QPOs, relative change in R$_{in}$ was noticed in one of the observations, although no consistent variation was seen in the power-law index except in a few observations.
Both flux and f$_{scat}$ were found to be increasing in two observations along with the disappearance of QPO, suggesting that corona or jet has increased in height.

6. A study of quasi-simulations X-ray and radio emission unveiled that jet is present during both type-C and type-B QPOs and a relatively stronger jet is present during type-C QPOs.
The high quality factor of type-C QPO suggests that these are not produced in the jet but instead occur at the base of the jet which is acting like a rigid inner flow/corona as discussed above.
Our studies suggest that type-B QPOs are caused by the precession of a weak jet with a small and weak corona at its base in the inner region of the accretion disk.
These results are similar to that reported by Kylafis et al. (2020) where a precessing jet model corroborated the type-B QPO,
however, more observations are indeed required to confirm this scenario.

The rapid transition of QPOs in BHXBs provide an opportunity to understand the dynamics of the Keplerian flow and jet/corona.
More studies are needed in order to investigate the contribution of the jet and ambient corona in modulating type-B and type-C QPOs in the inner region of the accretion disk.

\acknowledgments
We acknowledge the Referee for providing crucial inputs which improved the quality of the paper.
This work is based on observations made with {\it Rossi X-ray Timing Explorer} (RXTE).
This research has made use of data obtained through HEASARC Online Service, provided by
NASA/GSFC, in support of the NASA High Energy Astrophysics Programs. K.S. acknowledges the support from 
the SERB Core Research Grant, Government of India.
S.H. acknowledges the support from the SRF grant (Reg. No. 518980) from CSIR-UGC.
 Some part of the work has been carried out at
{\it Korea Astronomy and Space Science Institute}, Republic of Korea.

{\begin{table*}
\begin{minipage}[t]{\columnwidth}
\caption{Analyzed observations of the BH source H1743-322 }
\label{tab1}
\small
\begin{tabular}{ccccc}
\hline
\hline
ObsID & MJD (DD-MM-YY) & start time & end time & transition time (s) \\
\hline
 80146-01-07-00 & 52751.039 (22-04-2003) & 00:56:13 & 01:58:15 & 30 \\
 80146-01-13-00 & 52752.851 (23-04-2003) & 20:26:35 & 01:32:15 & 20\\
 80146-01-17-00 & 52756.186 (27-04-2003) & 04:28:13 & 06:03:15 & $>$2000\\
 80146-01-25-00 & 52763.075 (04-05-2003) & 01:48:57 & 02:47:15 & 40\\
 80146-01-50-00 & 52786.290 (27-05-2003) & 06:58:07 & 09:51:15 & 20 \\
 80135-02-02-000& 52787.280 (28-05-2003) & 06:44:01 & 14:25:08 & $~$280\\
 80146-01-51-01 & 52788.520 (29-05-2003) & 12:28:58 & 14:05:15 & 30 \\
 94413-01-03-04 & 54990.394 (08-06-2009) & 09:27:50 & 10:21:17 & 20\\
\hline
\end{tabular}
\end{minipage}
\end{table*}}

\begin{sidewaystable}
\flushleft
\caption{Log of the QPO parameters associated with the sudden transition observations. PDS power units is (rms/mean)$^{2}$ Hz$^{-1}$ and "weak" means QPO is weak, $\le$ 1 $\sigma$ with respect to the QPO normalization}
\label{tab1}
\tiny
\begin{tabular}{cc|cc|cc|cc|cc|cc|}

\hline
\hline
Parameter & Energy band (keV) & \hspace{0.5cm}80146-01-17-00 & & \hspace{0.5cm}80146-01-25-00 & & \hspace{0.5cm}80135-02-02-000 & & \hspace{0.5cm}80146-01-51-01 & & \hspace{0.5cm}94413-01-03-04 & \\

\hline
\hline

 &  & Q & NQ & Q & NQ & A & B & Q & NQ & Q & NQ\\
\hline
LC\footnote{Centroid frequency of the Lorentzian component} & 3.68--5.71 & 5.18$\pm$0.27\footnote{Fundamental centroid frequency}   &  no  &  no   & no  & 4.62$\pm$0.09(8.98$\pm$0.71) & 3.81$\pm$0.13(5.42$\pm$0.26) &   4.22$\pm$0.11(8.57$\pm$0.60)  & no & no & no \\

   & 6.12--14.76 & 5.12$\pm$0.06(9.44$\pm$0.40)\footnote{Harmonic centroid frequency} & 6.42$\pm$0.95(weak)  &  4.29$\pm$0.06(8.31$\pm$0.10) & 6.61$\pm$0.37   & 4.65$\pm$0.07(9.28$\pm$0.19) & 5.57$\pm$0.08 &  4.25$\pm$0.07(8.35$\pm$0.18)   &  no & 3.53$\pm$0.18 & no \\
\hline
LW\footnote{Width of the Lorentzian component} & 3.68--5.71 & 1.13$\pm$0.61   &  no  &  no  &   no  & 0.58$\pm$0.26(2.53$\pm$1.19) & 0.40$\pm$0.18(1.93$\pm$0.90) &  0.71$\pm$0.42(3.01$\pm$1.36)  & no & no & no \\

& 6.12--14.76 & 1.11$\pm$0.17(4.17$\pm$1.12)   &   4.43$\pm$2.01(weak)   &   0.90$\pm$0.11(2.29$\pm$0.47) & 5.86$\pm$0.95 & 0.97$\pm$0.19(2.04$\pm$0.39) & 1.57$\pm$0.16 & 1.16$\pm$0.20(3.71$\pm$0.52)  &  no & 1.42$\pm$0.50 & no \\
Q\footnote{$\frac{\nu}{\delta \nu}$, Quality factor of the fundamental frequency} &  3.68--5.71 & 4.58  & no    & no  & no & 7.97 (3.55)  & 9.52 (2.81) & 5.94 (2.85) & no & no & no \\
	&  6.12--14.76 & 4.56(2.26)  &  1.45  &  4.7(3.63)  &   1.13  & 4.79(4.55) & 3.55 &  3.66(2.25) & no & 2.48 & no \\
RMS (\%) & 3.68-5.71 &  2.59  &  no & no  & no & 2.07(1.23) & 1.95(2.76) & 1.60(1.60)  &  no & no & no \\
 &  6.12--14.76 & 4.87(3.61)  &  4.47  &  4.02(4.16)  &   4.90  & 5.32(3.27) & 7.01 &  3.41(4.43) & no & 5.13 & no \\
$\chi^{2}$/dof & 3.68--5.71 &  0.87 &  no & no  & no & 0.95 & 1.59  & 1.03  &  no & no & no \\
 &  6.12--14.76 & 0.86  &  1.35  &  0.95  &  1.19  & 1.02 & 1.41 &  1.17 & no & 0.92 & no \\
\hline
\hline
 Parameter &  & 80146-01-07-00 & & 80146-01-13-00 & & 80146-01-50-00 &  & & &&\\
\hline
\hline
  &  & Q & NQ & Q & NQ & Q & NQ  & & & &\\
 \hline
 LC & 3.68--5.71 & 9.49$\pm$0.27(weak)   &  no   &   no   &  no  &   9.38$\pm$0.06  &    no  & & & &\\
 & 6.12--14.76 & 9.70$\pm$0.10   &  no   &   9.66$\pm$0.05  &   no  &   9.45$\pm$0.03  &  no  & &  & &\\
\hline
  LW & 3.68--5.71 & 0.86$\pm$0.55(weak)   &   no   &  no   &  no   &   0.96$\pm$0.17  &    no  & & & &\\
  & 6.12--14.76 & 1.08$\pm$0.20   &   no   & 1.29$\pm$0.13  &  no  &   0.97$\pm$0.07  &  no  & & & &\\
Q &   3.68--5.71 & 11.03   &   no  & no  & no & 9.77 & no & & & & \\
  &  6.12--14.76 & 8.89   &  no   & 7.49   & no  &   9.74  & no & & & & \\
RMS (\%)&  3.68-5.71 & 1.92  & no   & no & no & 2.65 & no &  &   & &  \\
   & 6.12--14.76 & 4.73  &  no   & 5.22   & no   &   6.12  &  no & & & &\\
$\chi^{2}$/dof & 3.68--5.71 &  1.11  &  no & no  & no & 5.08  & no &   &  &  & \\
 &  6.12--14.76 & 1.12  &  no  &  1.23  &  no  & 1.21 & no & &  &  & \\
\hline
\hline
\end{tabular}
\end{sidewaystable}

\begin{sidewaystable}
\flushleft
\begin{minipage}[t]{\columnwidth}
\caption{Best-fit spectral parameters of the different sections (Q/NQ \& A/B) of the observations exhibiting QPO $\nu$ $\sim$ 4.5 Hz  \& 3.5 Hz.}
\label{tab1}
\tiny
\begin{tabular}{c|cc|cc|cc|cc|cc}
\hline
\hline
Sections & 80146-01-17-00 & & 80146-01-25-00  & & 80135-02-02-000 & & 80146-01-51-01  & & 94413-01-03-04 \\
\hline
\hline
Parameter & Q & NQ & Q & NQ & A & B & Q & NQ & Q & NQ \\
\hline
 & & &  & & model 1 &  & &  & \\
\hline
$kT_{in}$ (keV)\footnote{Inner disk temperature of the diskbb model}  &  1.31$\pm$0.01   &   1.24$\pm$0.01   &   1.23$\pm$0.01    &  1.19$\pm$0.01  & 1.21$\pm$0.01   &   1.24$\pm$0.01  &   1.17$\pm$0.01   &   1.17$\pm$0.01 &   0.72$\pm$0.03   &   0.76$\pm$0.03\\
N$_{diskbb}$\footnote{Normalization of the diskbb model}   &   332$\pm$12   &   471$\pm$10   &   232$\pm$7   &   273$\pm$5  &   639$\pm$17   &   529$\pm$22 &   731$\pm$15    &    849$\pm$20 &   1089$\pm$234   &   909$\pm$151 \\
$\Gamma$\footnote{Power-law index}     &   2.79$\pm$0.03   &   2.70$\pm$0.04   &   2.83$\pm$0.03   &   2.69$\pm$0.03 &   2.65$\pm$0.04   &   2.65$\pm$0.04  &   2.61$\pm$0.03  &    2.41$\pm$0.05 &   2.50$\pm$0.07   &   2.32$\pm$0.08  \\
N$_{pl}$\footnote{Normalization of the power-law
model (Photons  $keV^{-1} cm^{-2} s^{-1} at 1 keV$)}   &  10.65$\pm$1.01   &   5.36$\pm$0.61   &   4.63$\pm$0.48   &   2.57$\pm$0.25  &  8.08$\pm$0.88   &   12.17$\pm$1.34  &   5.56$\pm$0.49  &    3.25$\pm$0.48  &  1.80$\pm$0.41   &   1.09$\pm$0.28 \\
diskbb flux\footnote{Unabsorbed flux} for all the models is in units $10^{-9}$ erg cm$^{-2} s^{-1}$ in the energy band 3-25 keV   &  8.01   &   8.51  &   3.99  &   4.05 &   10.20  &    9.68 &   10.01  &    11.12  &   0.77  &    0.93 \\
power-law flux &  7.34   &   4.37  &   2.97  &   2.14  &   7.27  &   10.98 &   5.58  &    4.71 &   2.17  &   1.91\\
$\chi^{2}$/dof  &  48/44   &   33/44   &   51/44   &   52/44   &   39/44   &   40/44 &   27/44  &    35/44  &   62/42   &   72/42\\
$\nu\footnote{QPO centroid frequency} /Q_{\nu}=\nu/\delta \nu\footnote{Quality factor}$  &  5.12$\pm$0.06/4.57  &   -   &   4.29$\pm$0.06$^{h}$/4.76   &   - &  4.65$\pm$07/5.44  &   5.57$\pm$0.08/3.54  &   4.25$\pm$0.07/3.66  &    - &  3.46$\pm$0.22/3.62  &   -  \\


R$_{in}\footnote{Inner accretion disk radius}$(km) & 37$\pm$1 & 43$\pm$1 & 31$\pm$1 & 33$\pm$1  & 51$\pm$1 & 46$\pm$1 & 54$\pm$1 & 58$\pm$1  & 66$\pm$7 & 60$\pm$5 \\
\hline
 & & &  & & model 2 &  & &  & \\
\hline
$kT_{in}$ (keV)  & 1.29$\pm$0.02 & 1.23$\pm$0.01 & 1.22$\pm$0.01 & 1.21$\pm$0.03  & 1.21$\pm$0.01 & 1.21$\pm$0.01  & 1.17$\pm$0.01 & 1.17$\pm$0.01  & 0.71$\pm$0.03 & 0.69$\pm$0.04 \\
N$_{diskbb}$ & 374$\pm$44 & 513$\pm$40 & 248$\pm$35 & 235$\pm$32 & 689$\pm$40 & 688$\pm$58 & 736$\pm$21 & 854$\pm$23  & 1136$\pm$119 & 1172$\pm$286 \\
$\Gamma_{1}$\footnote{$\Gamma_{1} \& \Gamma_{2}$ are the photon indices of broken-power-law model} & 2.66$\pm$0.03 & 2.48$\pm$0.04 & 2.72$\pm$0.02 & 2.93$\pm$0.03 & 2.48$\pm$0.01 & 2.33$\pm$0.02 & 2.55$\pm$0.08 & 2.36$\pm$0.03  & 2.55$\pm$0.01 & 2.64$\pm$0.03  \\
E$_{b}$\footnote{Break energy of the broken power-law} & 11.80$\pm$2.42 & 10.79$\pm$1.71 & 13.66$\pm$2.32 & 9.60$\pm$1.88 & 11.69$\pm$1.77 & 9.71$\pm$1.30 & 16.18$\pm$1.83 & 13.91$\pm$1.62 & 17.96$\pm$2.85 & 13.33$\pm$1.45 \\
$\Gamma_{2}$  & 2.81$\pm$0.05 & 2.71$\pm$0.05 & 2.92$\pm$0.09 & 2.68$\pm$0.048 & 2.67$\pm$0.05 & 2.66$\pm$0.04 & 2.67$\pm$0.06 & 2.46$\pm$0.06  & 2.16$\pm$0.53 & 2.16$\pm$0.20 \\
N$_{bkn}$\footnote{Normalization (photons $cm^{-2} s^{-1} keV^{-1}$  at  1 keV)} & 7.80$\pm$1.02 & 3.14$\pm$1.68 & 3.52$\pm$0.88 & 4.51$\pm$2.02  & 5.31$\pm$1.98 & 5.86$\pm$3.34 & 4.98$\pm$1.04 & 2.62$\pm$0.02  & 2.03$\pm$0.67 & 2.42$\pm$0.82 \\
diskbb flux   &  8.52   &   9.04  &   4.23   &   3.78   &  10.87  &    11.10  &   10.03  &    11.02 &   0.72  &    0.62\\
bknpower-law flux &  6.81   &   3.87  &   2.75   &   2.42   &  6.58 &    9.51  &   5.45  &    4.47 &  2.23 &    2.29\\
$\chi^{2}$/dof  &  43/42   &   32/42   &   44/42   &   47.6/42    &   34/42 &    33/42 &   24/42  &    33.8/42  &   51/42 &    55/42\\
R$_{in}$(km) & 39$\pm$3 & 45$\pm$2 & 32$\pm$1 & 31$\pm$2  & 53$\pm$2 & 53$\pm$2 & 54$\pm$2 & 59$\pm$2  & 68$\pm$4 & 69$\pm$8\\
\hline
\end{tabular}
\end{minipage}
\end{sidewaystable}


{\begin{table*}
\begin{minipage}[t]{\columnwidth}
\caption{Best-fit spectral parameters of the different sections (Q and NQ) of the observations exhibiting QPO $\nu$ $\sim$ 9.5 Hz .}
\label{tab1}
\tiny
\begin{tabular}{ccccccc}
\hline
\hline
Sections & 80146-01-07-00 & & 80146-01-13-00 & & 80146-01-50-00\\
\hline
\hline
Parameter & Q & NQ & Q & NQ & Q & NQ\\
\hline
 & & & model 3 & &  & \\
\hline
$kT_{in}$ (keV)  &  1.21$\pm$0.01   &   1.25$\pm$0.01   &   1.32$\pm$0.01    &    1.47$\pm$0.01   &   1.25$\pm$0.01   &   1.36$\pm$0.02\\
N$_{diskbb}$   &   137$\pm$12   &   105$\pm$14   &   287$\pm$20   &   182$\pm$17   &   429$\pm$14  &  293$\pm$29\\
kT$_{s}$\footnote{Input soft seed photon temperature}     &   0.33   &   0.27   &   0.23  &   0.23   &   0.41  &    0.52\\
kT$_{e}$\footnote{Electron temperature}(KeV) &  10.14$\pm$0.58   &   7.47$\pm$1.01   &   7.83$\pm$0.91   &   6.32$\pm$0.38   &   14.12$\pm$0.18  &    9.47$\pm$1.66\\
$\tau$\footnote{Optical depth}   &  3.89$\pm$0.26   &   5.25$\pm$0.62  &   5.37$\pm$0.57   &   6.61$\pm$0.41   &   3.51$\pm$0.52  &    4.83$\pm$0.65\\
N$_{comptt}$ &   0.55$\pm$0.05   &   0.79$\pm$0.06   &   1.71$\pm$0.08   &   2.57$\pm$0.12   &   0.53$\pm$0.02    &   0.71$\pm$0.04\\
diskbb flux\footnote{Unabsorbed flux in the energy range 3.0-25.0 keV (10$^{-9}$ erg cm$^{-2}$ s$^{-1}$)}   &  2.19   &   2.04  &   7.41   &   8.10   &   8.39  &    8.91 \\
CompTT flux &  6.61   &   6.03  &   12.90   &   17.80   &   14.80  &    18.66 \\
$\chi^{2}$/dof  &  58/43   &   56/43   &   45/43   &   37/43  &   29/43  &    33/43\\
$\nu /Q_{\nu}$  &  9.7$\pm$0.10/8.89  &   -   &   9.66$\pm$0.05/7.49   &   -   &   9.45$\pm$0.03/9.74  & - \\
R$_{in}$(km) & 23$\pm$2 & 20$\pm$2 & 34$\pm$2 & 27$\pm$1 & 42$\pm$2 & 35$\pm$2 \\
\hline
 & & & model 2 & &  & \\
\hline
$kT_{in}$ (keV)  & 1.24$\pm$0.03 & 1.32$\pm$0.03 & 1.36$\pm$0.01 & 1.54$\pm$0.02 & 1.27$\pm$0.01 & 1.38$\pm$0.03 \\
N$_{diskbb}$ & 118$\pm$25 & 81$\pm$18 & 267$\pm$24 & 156$\pm$17 & 390$\pm$38 & 310$\pm$48 \\
$\Gamma_{1}$ & 2.73$\pm$0.06 & 2.62$\pm$0.07 & 2.45$\pm$0.06 & 2.37$\pm$0.05 & 2.51$\pm$0.07 & 2.36$\pm$0.02 \\
E$_{break}$ & 12.28$\pm$1.03 & 12.95$\pm$1.01 & 13.16$\pm$0.47 & 13.88$\pm$0.41 & 13.04$\pm$1.93 & 12.49$\pm$0.88 \\
$\Gamma_{2}$  & 3.02$\pm$0.05 & 3.01$\pm$0.06 & 2.89$\pm$0.03 & 2.94$\pm$0.03 & 2.67$\pm$0.03 & 2.72$\pm$0.05 \\
N$_{bkn}$(km) & 8.72$\pm$1.28 & 6.35$\pm$1.01 & 9.47$\pm$1.55 & 11.07$\pm$1.42 & 12.91$\pm$2.19 & 11.11$\pm$2.77 \\
diskbb flux   &  2.19   &   2.08  &   8.01   &   8.89   &   8.02  &    9.91\\
bknpower-law flux &  6.60   &   5.98  &   12.65   &   16.71   &   15.20  &    17.01\\
$\chi^{2}$/dof  &  49/40   &   63/40   &   34.4/42   &   39/42  &   26.5/40  &    29/40\\
R$_{in}$(km) & 22$\pm$2 & 18$\pm$2 & 33$\pm$2 & 25$\pm$2 & 40$\pm$2 & 35$\pm$2 \\
\hline
\end{tabular}
\end{minipage}
\end{table*}}

{\begin{table*}
\begin{minipage}[t]{\columnwidth}
\caption{Best-fit spectral parameters of the sections with QPO \& non-QPO using the model wabs$\times$(SIMPL*diskbb).}
\label{tab1}
\tiny
\begin{tabular}{c|cc|cc|cc|cc|cc|}
\hline
\hline
 & 80146-01-17-00  & &80146-01-25-00  & & 80135-02-02-000 & & 80146-01-51-01 & & 94413-01-03-04\\
\hline
\hline
Parameter & Q & NQ & Q & NQ & A & B & Q & NQ & Q & NQ \\
\hline
\hline
kT$_{in}$\footnote{Inner disk temperature (keV)}     &  1.20$\pm$0.02   &  1.19$\pm$0.01   &   1.15$\pm$0.01  & 1.16$\pm$0.01   & 1.15$\pm$0.01 & 1.13$\pm$0.01 & 1.14$\pm$0.01  &    1.15$\pm$0.01  &  0.96$\pm$0.04   &  0.98$\pm$0.03\\
N$_{dbb}$\footnote{Normalization} &  790$\pm$52   &   738$\pm$50  &   459$\pm$26   &   409$\pm$16 & 1135$\pm$62 & 1322$\pm$90  &   1092$\pm$42  &   1106$\pm$52   & 448$\pm$90   &   412$\pm$74\\
$\Gamma_{simpl}$\footnote{Power-law index of the SIMPL model}  &  2.91$\pm$0.07   &   2.76$\pm$0.06   &   2.92$\pm$0.05    &   2.68$\pm$0.05 & 2.71$\pm$0.06 & 2.74$\pm$0.06  &  2.60$\pm$0.04   &   2.38$\pm$0.07  &  2.40$\pm$0.06   &   2.32$\pm$0.05 \\
f$_{scat}$\footnote{Scattered fraction of the soft photons} &   0.121$\pm$0.006   &  0.193$\pm$0.010  &   0.161$\pm$0.008   &  0.15$\pm$0.004  & 0.156$\pm$0.008 & 0.223$\pm$0.008  &   0.124$\pm$0.007    &    0.101$\pm$0.005  &   0.274$\pm$0.015   &  0.262$\pm$0.013\\
$\chi^{2}$/dof  &  43/42   &   38/42   &   44/42   &   47/42   & 35/42 & 36/42 &   24/42  &    32/42  &   54/42  &    53/4\\
R$_{in}$(km) & 56$\pm$3   & 54$\pm$3   & 43$\pm$2   & 40$\pm$1 & 67$\pm$4 & 72$\pm$5 & 66$\pm$2  & 66$\pm$3  & 42$\pm$8  & 40$\pm$7 \\
flux\footnote{Unabsorbed flux in the energy range 3.0-25.0 keV (10$^{-9}$ erg cm$^{-2}$ s$^{-1}$)}& 15.26& 12.87 & 6.94 & 6.18 & 17.41&20.54&15.51 &15.82& 2.97&2.95\\
\hline
   & 80146-01-07-00 & & 80146-01-13-00 & & 80146-01-50-00\\
\hline
\hline
Parameter & Q & NQ & Q & NQ & Q & NQ   \\
\hline
kT$_{in}$ &   0.94$\pm$0.03   &   0.93$\pm$0.01   &  1.10$\pm$0.02  &  1.02$\pm$0.02   & 1.12$\pm$0.01  & 1.12$\pm$0.03\\
N$_{dbb}$ &  1221$\pm$92   & 1405$\pm$110  &  1338$\pm$120  &  2160$\pm$195   &   1448$\pm$100  & 1607$\pm$250\\
$\Gamma_{simpl}$ &  3.02$\pm$0.01   &  3.05$\pm$0.01   &   2.92$\pm$0.02    &  3.00$\pm$0.02   &  2.67$\pm$0.02   &  2.78$\pm$0.04\\
f$_{scat}$ &   0.44$\pm$0.04   &   0.53$\pm$0.08   &   0.38$\pm$0.02   &    0.55$\pm$0.06   &  0.32$\pm$0.01   &  0.40$\pm$0.02\\
$\chi^{2}$/dof  &  52/42   &   51/42   &   35/42   &   40/42   &   34/42  &    34/42\\
R$_{in}$(km) & 70$\pm$5  & 75$\pm$5  & 73$\pm$6  & 93$\pm$8  & 76$\pm$5  & 80$\pm$12 \\
flux& 8.81& 10.10 & 20.31 & 25.80 & 21.42&27.33\\
\hline
\hline
\end{tabular}
\end{minipage}
\end{table*}}

\clearpage
{\begin{table*}
\begin{minipage}[t]{\columnwidth}
\caption{Simultaneous spectral fit results with two different models along with F-test probabilities for two sample observations. All fixed means for example for ObsID 80146-01-17-00, the spectral parameters of Q and NQ sections are tied and the parameters are fixed as the best-fit values of NQ section spectrum. Each model parameter were allowed to vary independently by freeing in Q and untying in NQ sections.}
\label{tab1}
\tiny
 \begin{tabular}{ccc}
 \hline
\hline
      Parameters & $\chi^2$/dof &  F-test probability \\ 
        \hline
	&  80135-02-02-000 &wabs$\times$(SIMPL*diskbb)	 \\ 
         \hline
        All fixed & 12860/92 & --  \\
       $\Gamma_{simpl}$ & 8446/90 &  5.94$\times$ 10$^{-9}$ \\
        f$_{scat}$ & 611/88 & 6.54$\times$ 10$^{-51}$ \\
        kT$_{in}$ & 78/86 & 5.94$\times$ 10$^{-39}$ \\
        All freed & 70/84 &  6.50$\times$ 10$^{-3}$ \\   \hline
          &  80146-01-17-00	& model 2	 \\ \hline
            All fixed & 36032/96 & -  \\
	kT$_{in}$ & 16899/94 &  3.50$\times$ 10$^{-16}$ \\
	N$_{diskbb}$ & 12184/92 &  2.91$\times$ 10$^{-7}$ \\
       $\Gamma_{1}$ & 101/90 &  2.15$\times$ 10$^{-94}$ \\
	E$_{b}$ & 100/88 &  0.65 \\
	All freed & 75/86 &  4.20$\times$ 10$^{-3}$ \\ \hline   
      \end{tabular}
    \end{minipage}
\end{table*}

{\begin{table*}
\begin{minipage}[t]{\columnwidth}
\small
\caption{Log of X-ray and radio observations associated with a sudden transition. Radio fluxes were taken from McClintock et al (2009) and Miller-Jones et al. (2012).}
\label{tab5}
\tiny
\begin{tabular}{cccccc}
\hline
\hline
ObsID & X-ray MJD (QPO type)& Radio MJD (nearest) & Radio flux (4.860 GHz, mJy)& Radio flux (8.460 GHz, mJy)& Time difference\\
\hline
\hline

80146-01-17-00& 52756.219 (B)& 52755.531 $< >$ 52758.421& 7.54 $< >$ 8.45& 5.58 $<>$ 6.64&More than a day \\
80146-01-25-00& 52763.095 (B)& 52765.437&13.78 & 11.12&More than a day\\
80135-02-02-000 &52787.237 (B)& 52786.484& 22.49 (MJD 52786.348)& 16.15--16.90 & $\sim$ 18 h\\
80146-01-51-01&52788.506 (B)& 52788.375& NA&5.59& $\sim$3.2 h\\
94413-01-03-04& 54990.26 (A)&54990.33& --&$<$1.0&$\sim$1.7 h\\
\hline
80146-01-07-00& 52751.089(C)&52750.597$<>$52752.535 & 23.10$<>$12.20&20.94$<>$11.48 &More than a day\\
80146-01-13-00& 52752.972(C)& 52752.535& 12.20& 11.48& $\sim$10.50 h\\
80146-01-50-00& 52786.350(C)& 52786.355&22.49&15.81&0.12 h\\
\hline
\end{tabular}
\end{minipage}
\end{table*}}
\clearpage
\begin{figure*}
\centering
\includegraphics[height=20cm,width=20cm,angle=-90]{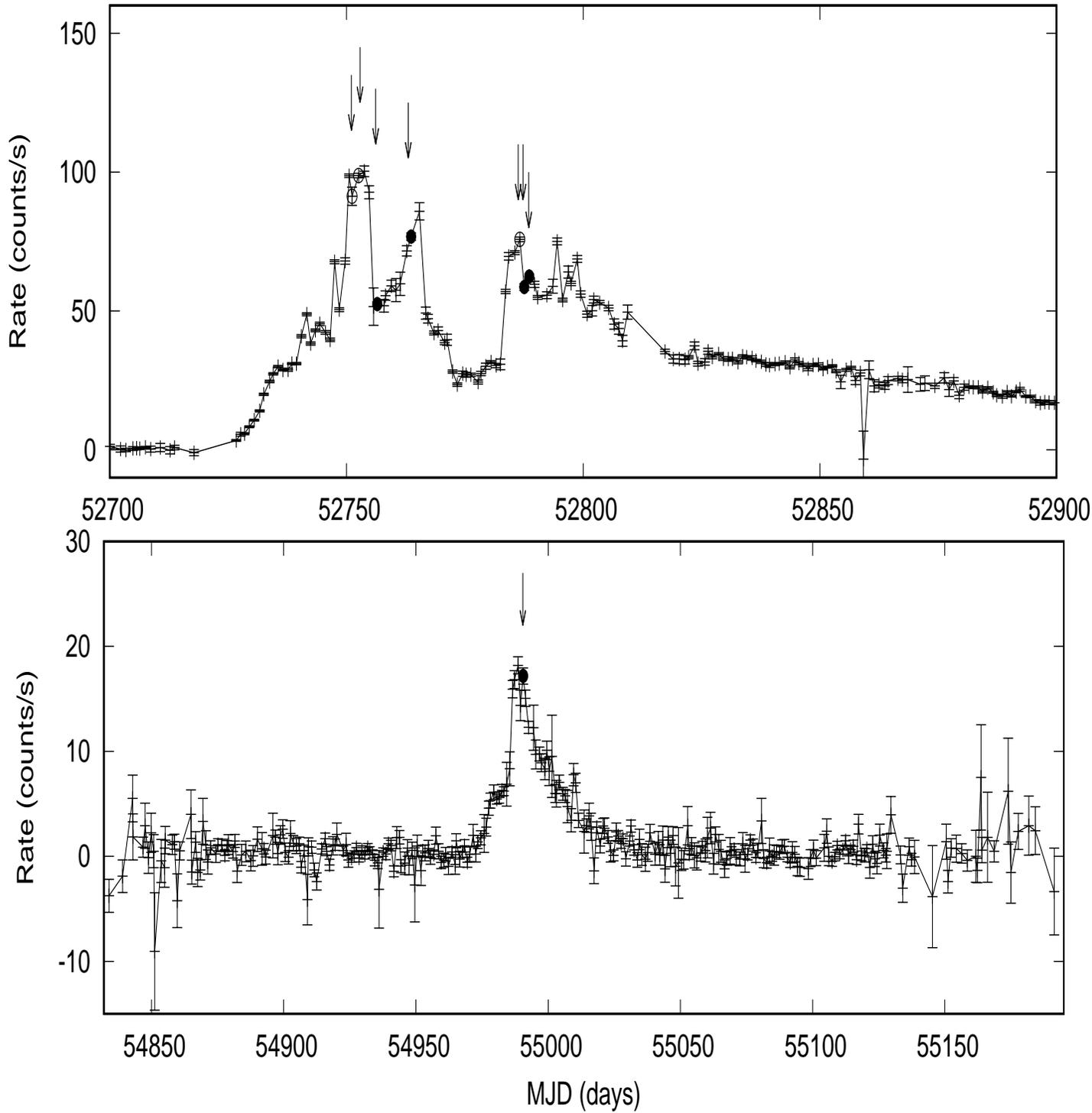}
\caption{ASM light curves indicating rapid transition events for 2003 outburst (upper panel),
open circle shows the type-C QPOs and filled circle shows the type-B QPOs  associated events, and 2009 outburst (lower panel) with filled circle shows the type-A QPO event}.
       \label{Fig1}
 \end{figure*}

\clearpage
\begin{figure*}
\centering
\includegraphics[height=20cm,width=20cm,angle=-90]{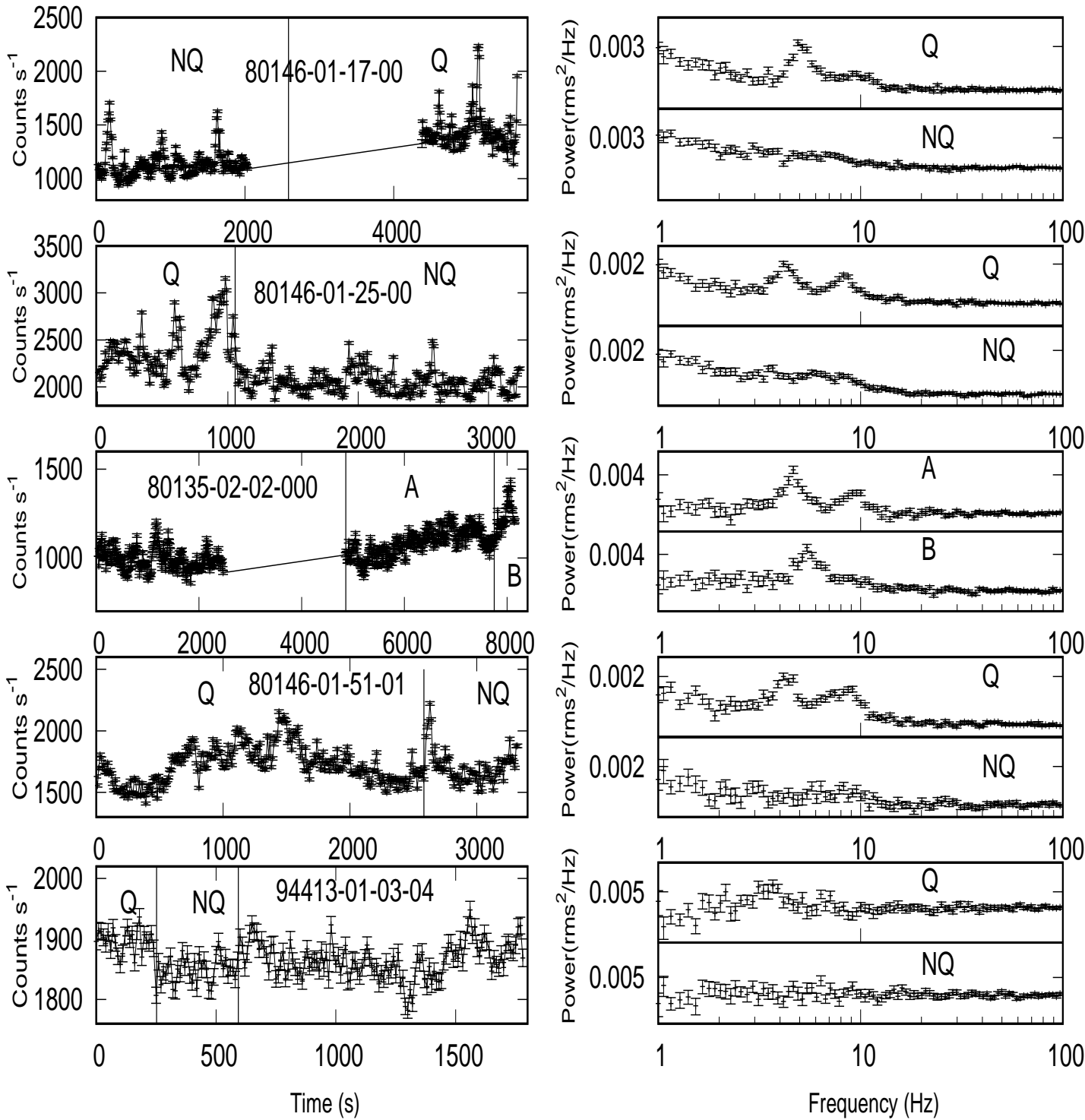}
\caption{Left: light curves in 6.12--14.76 keV energy band with a bin time of 8 s.
QPO (Q) and non-QPO (NQ) sections are shown with Q and NQ labels, segregation is marked with a vertical line.
QPOs of the ObsID 80135-02-02-00 data are not disappeared and transition is shown with A and B sections.
Right: respective PDS of QPO (Q) and non-QPO (NQ) sections. }

       \label{Fig2}
 \end{figure*}

\clearpage
\begin{figure*}
\centering
\includegraphics[height=20cm,width=20cm,angle=-90]{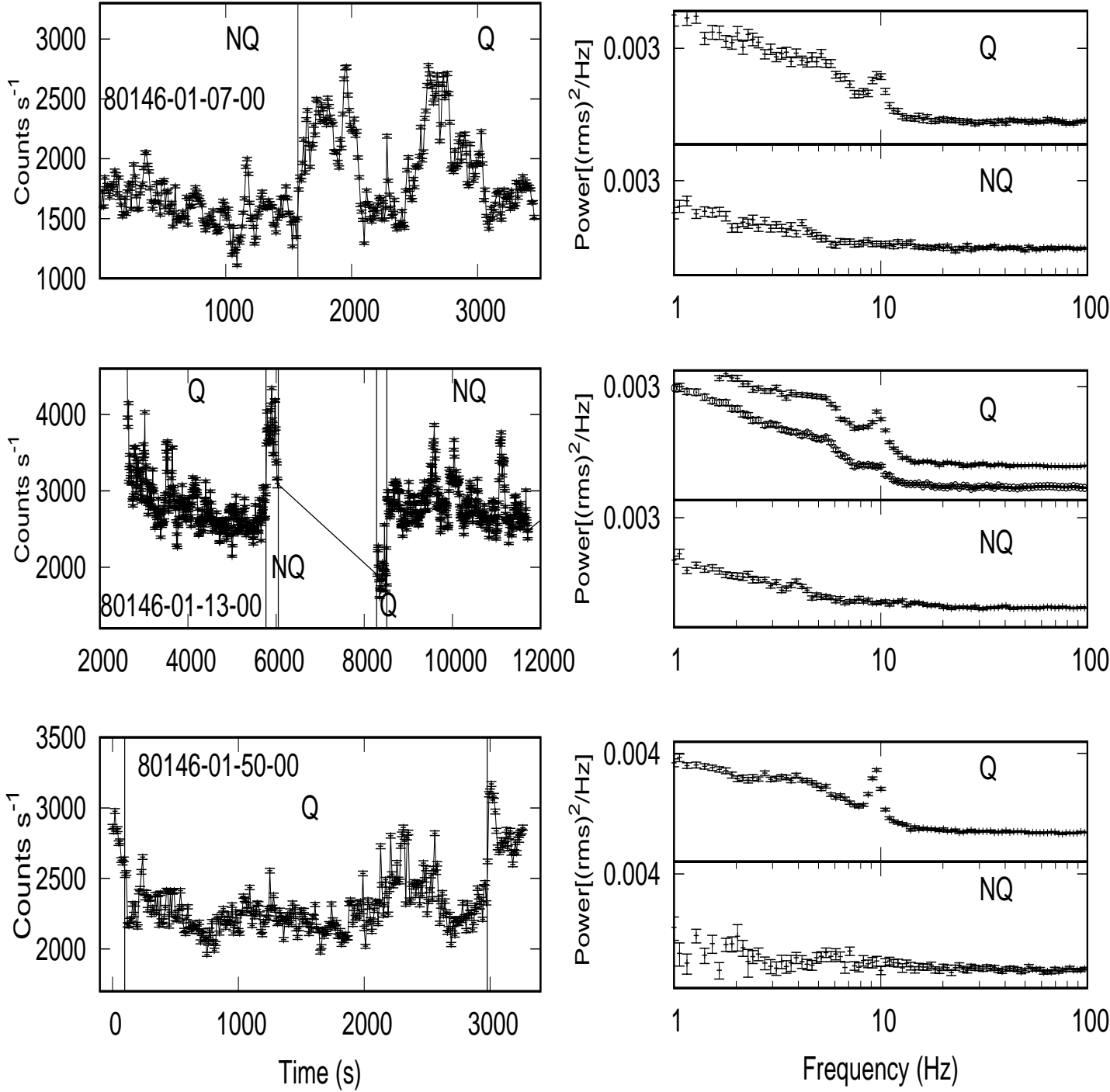}
\caption{Same as Figure 2. except for the PDS shown in circle 
for ObsID 80146-01-13-00 where QPO is absent in the lower energy band (3.68--5.71 keV) when compared to the PDS of higher band (6.12--14.76 keV).}
       \label{Fig3}
 \end{figure*}

\begin{figure*}
\begin{minipage}{0.5\textwidth}
\includegraphics[height=8.6cm,width=4.8cm,angle=-90]{1700spec_Q.ps}
\end{minipage}
\begin{minipage}{0.5\textwidth}
\includegraphics[height=8.6cm,width=4.8cm,angle=-90]{1700spec_NQ.ps}
\end{minipage}
\begin{minipage}{0.5\textwidth}
\includegraphics[height=8.6cm,width=4.8cm,angle=-90]{2500spec_Q.ps}
\end{minipage}
\begin{minipage}{0.5\textwidth}
\includegraphics[height=8.6cm,width=4.8cm,angle=-90]{2500spec_NQ.ps}
\end{minipage}

\begin{minipage}{0.5\textwidth}
\includegraphics[height=8.6cm,width=4.8cm,angle=-90]{000spec_A.ps}
\end{minipage}
\begin{minipage}{0.5\textwidth}
\includegraphics[height=8.6cm,width=4.8cm,angle=-90]{000spec_Bp.ps}
\end{minipage}

\begin{minipage}{0.5\textwidth}
\includegraphics[height=8.6cm,width=4.8cm,angle=-90]{5101spec_Q.ps}
\end{minipage}
\begin{minipage}{0.5\textwidth}
\includegraphics[height=8.6cm,width=4.8cm,angle=-90]{5101spec_NQ.ps}
\end{minipage}

\begin{minipage}{0.5\textwidth}
\includegraphics[height=8.6cm,width=4.8cm,angle=-90]{94413spec_Q1.ps}
\end{minipage}
\begin{minipage}{0.5\textwidth}
\includegraphics[height=8.6cm,width=4.8cm,angle=-90]{94413spec_NQ1.ps}
\end{minipage}
\caption{ Unfolded Spectra of QPO(Q)/non-QPO(NQ) \& A/B sections where type-B \& type-A QPOs were observed. Dashed lines show the model components ( model 1).}
       \label{}
 \end{figure*}

\clearpage
\begin{figure*}
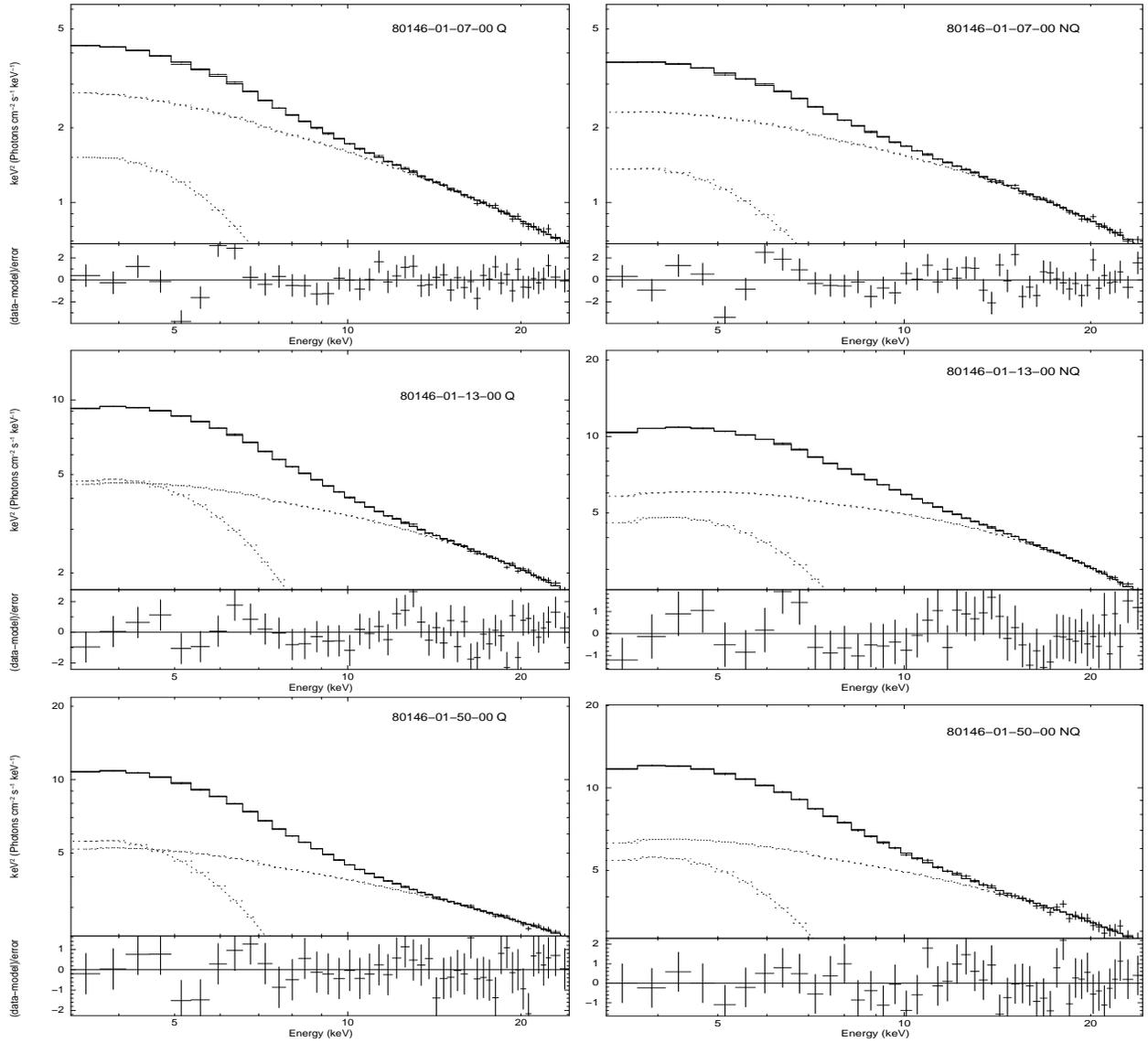

\begin{minipage}{0.5\textwidth}
\includegraphics[height=8.6cm,width=5cm,angle=-90]{0700spec_Q.ps}
\end{minipage}
\begin{minipage}{0.5\textwidth}
\includegraphics[height=8.6cm,width=5cm,angle=-90]{0700spec_NQ.ps}
\end{minipage}

\begin{minipage}{0.5\textwidth}
\includegraphics[height=8.6cm,width=5cm,angle=-90]{1300spec_Q.ps}
\end{minipage}
\begin{minipage}{0.5\textwidth}
\includegraphics[height=8.6cm,width=5cm,angle=-90]{1300spec_NQ.ps}
\end{minipage}

\begin{minipage}{0.5\textwidth}
\includegraphics[height=8.6cm,width=5cm,angle=-90]{5000spec_Q.ps}
\end{minipage}
\begin{minipage}{0.5\textwidth}
\includegraphics[height=8.6cm,width=5cm,angle=-90]{5000spec_NQ.ps}
\end{minipage}

\caption{ Unfolded spectra of QPO (Q) and non-QPO (NQ) section for type-C QPO observations. Dashed line show the model components ( model 3)}
       \label{Fig3}
 \end{figure*}
\begin{figure*}
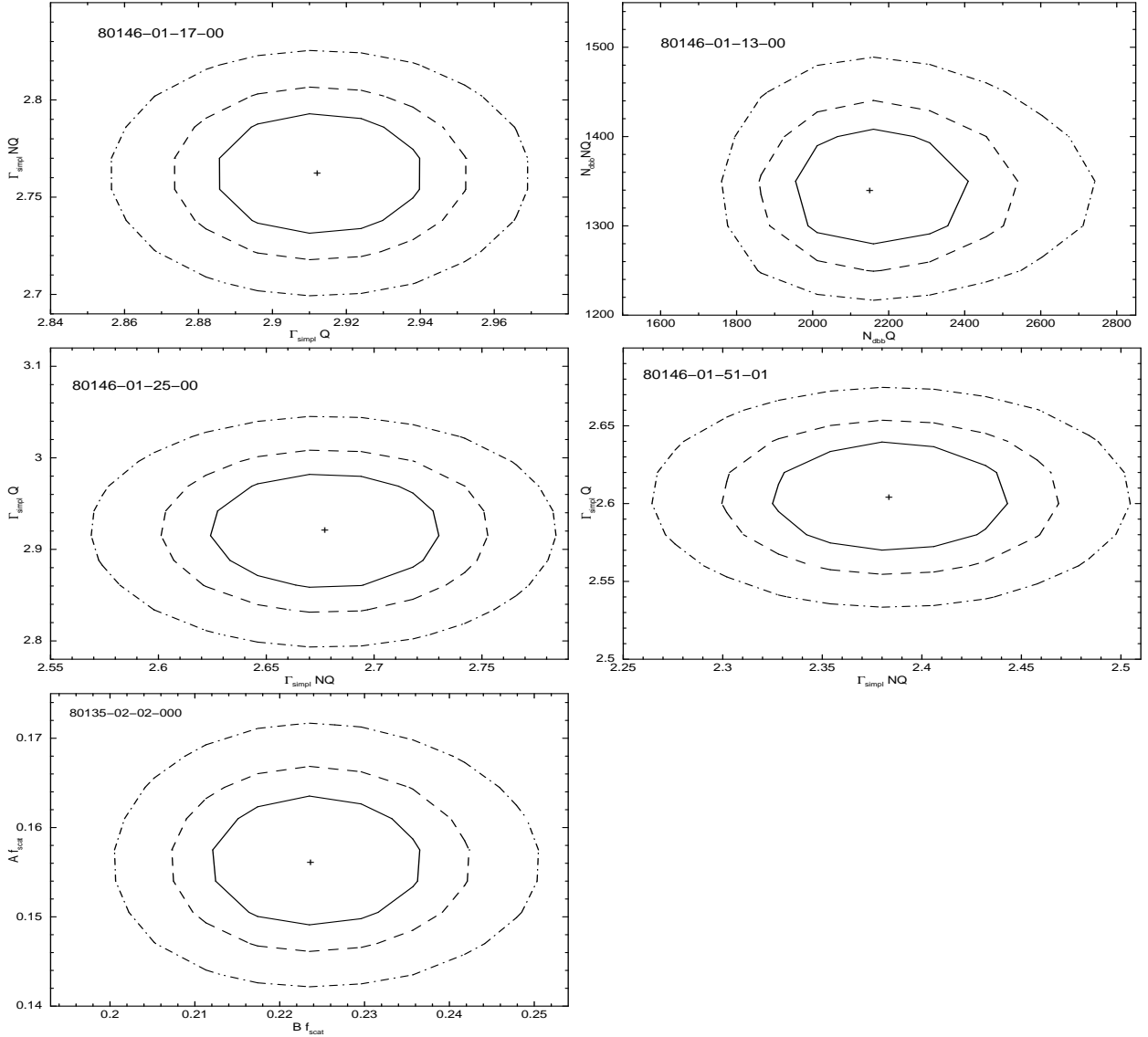

\begin{minipage}{0.5\textwidth} 
\includegraphics[height=8.6cm,width=5cm,angle=-90]{cont_17.ps}
\end{minipage}
\begin{minipage}{0.5\textwidth}
\includegraphics[height=8.6cm,width=5cm,angle=-90]{cont_13.ps}
\end{minipage}
\begin{minipage}{0.5\textwidth} 
\includegraphics[height=8.6cm,width=5cm,angle=-90]{cont_25.ps}
\end{minipage}
\begin{minipage}{0.5\textwidth}
\includegraphics[height=8.6cm,width=5cm,angle=-90]{cont_51new.ps}
\end{minipage}
\begin{minipage}{0.5\textwidth} 
\includegraphics[height=8.6cm,width=5cm,angle=-90]{cont_000.ps}
\end{minipage}
\caption{For each panel, confidence contours represent the uncertainty ranges, i.e. 1 $\sigma$ inner contour (line), 2 $\sigma$ middle contour (dashed line), and 3 $\sigma$ outer contour (dotted-dashed line). Each panel has the respective ObsIDs and the X \& Y axis represent the spectral parameters used obtaining the confidence contours. Each panel clearly shows that the respective spectral parameter varied between different sections (Q vs NQ and A vs B) and the variations are significant at the 99\% confidence level.} 
       \label{}
 \end{figure*}

\clearpage
\begin{figure*}
\begin{minipage}{0.5\textwidth} 
\includegraphics[height=8.6cm,width=9cm,angle=-90]{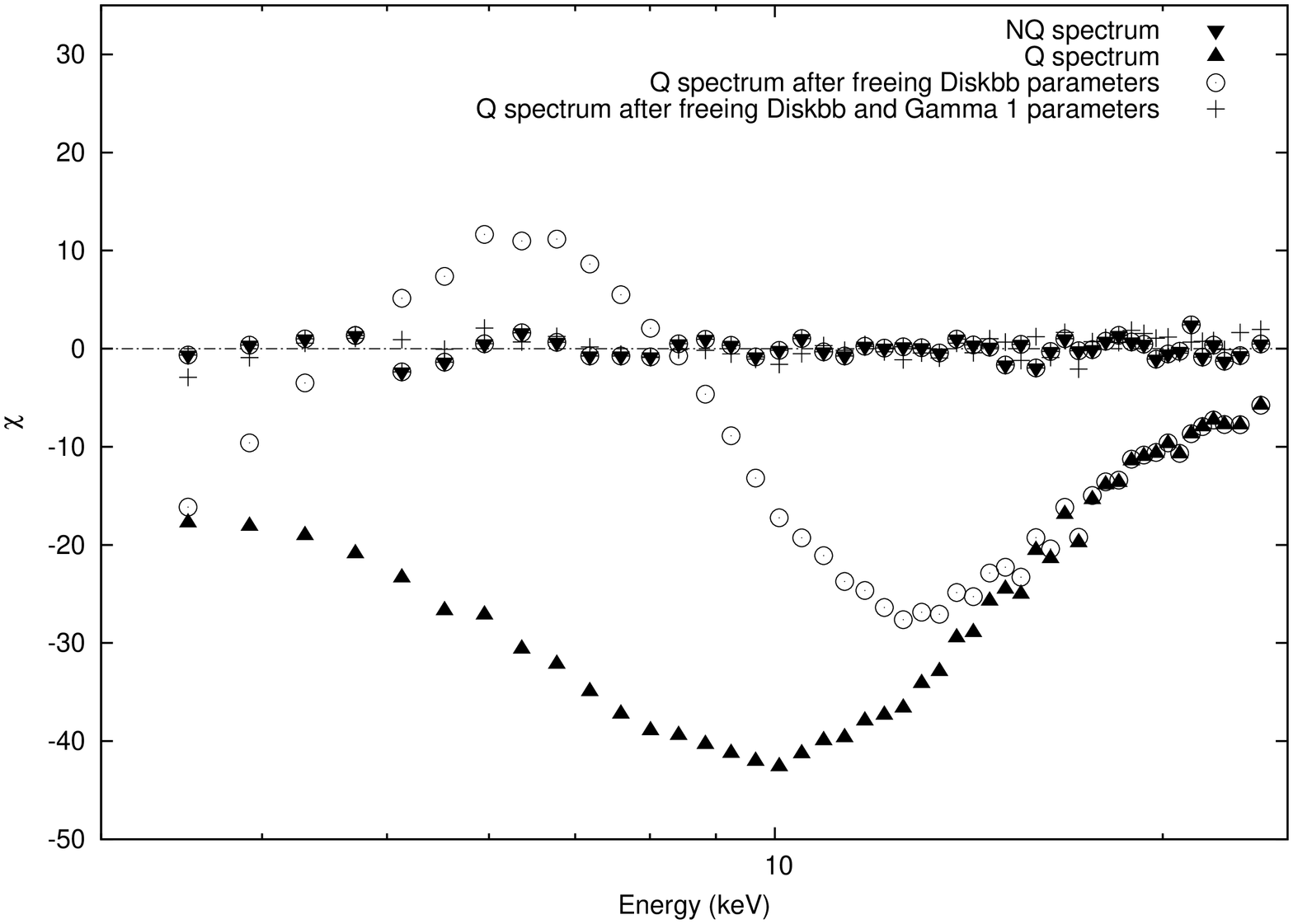}
\end{minipage}
\begin{minipage}{0.5\textwidth} 
\includegraphics[height=8.6cm,width=9cm,angle=-90]{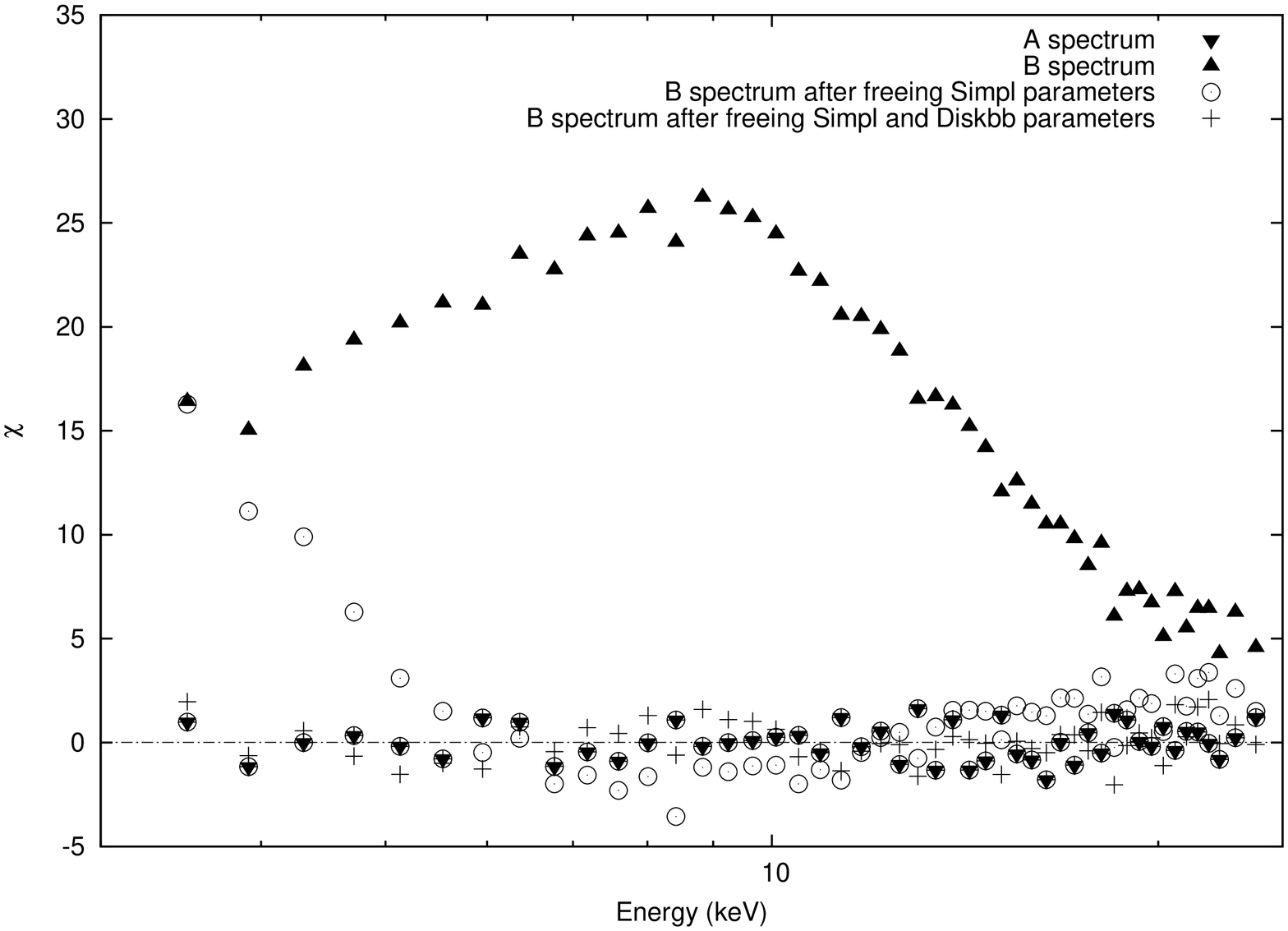}
\end{minipage}
\caption{Left: square root of $\chi^2$ for the spectra of Q and NQ sections of ObsID 80146-01-17-00. The best-fit NQ spectrum is close along the zero residual (inverted triangle) and as the spectral parameters of the model components are freed and untied (see text), the residuals of Q spectrum decreased.  Right: similar study for the spectra of A and B sections of ObsID 80135-02-02-000.} 
       \label{}
 \end{figure*}



\clearpage
\begin{thebibliography}{}
\small
\bibitem[Arnaud 1996]{ar96}Arnaud, K. A. 1996, in ASP Conf. Ser. 105, Astronomical Data Analysis
Software and Systems V, ed. G. H. Jacoby \& J. Barnes (San, Francisco,
CA: ASP), 17
\bibitem[]{}Arur, K., \& Maccarone, T. J. 2019, MNRAS, 486, 3451
\bibitem[]{}Arur, K., \& Maccarone, T. J. 2020, MNRAS, 491, 313
\bibitem[]{}Axelsson, M., \& Done, C. 2016, MNRAS, 458, 1778
\bibitem[]{}Axelsson, M., Hjalmarsdotter, L., \& Done, C. 2013, MNRAS, 431, 1987
\bibitem[]{}Baba, D., Nagata, T., Iwata, I., Kato, T., \& Yamaoka, H. 2003, IAUC, 8112, 2
\bibitem[]{}Belloni, T., Homan, J., Yamaoka, K., \& Swank, J. 2008, ATel, 1804, 1
\bibitem[]{}Belloni, T. M. 2010, in The Jet Paradigm: From Microquasars to Quasars, ed.
T. Belloni (Lecture Notes in Physics, Vol. 794; Berlin: Springer), 53
\bibitem[]{}Belloni, T. M., Motta, S. E., \& Munoz-Darias, T. 2011, BASI, 39, 409
\bibitem[]{}Belloni, T. M., \& Motta, S. E. 2016, Transient black hole binaries, in Astrophysics of Black Holes: From Fundamental Aspects to Latest Developments, ed. by C. Bambi (Springer, Berlin, 2016). arXiv:1603.07872 [astro-ph.HE]
\bibitem[]{}Belloni, T. M., Zhang, L., Kylafis, N, D., Reig, P., \& Altamirano, D. 2020, MNRAS, 496, 4366
\bibitem[]{}Beloborodov, A. 1999, ApJL, 510, L123
\bibitem[]{}Bogensberger, D., Ponti, G., Jin, C., et al. 2020, A\&A, accepted, [arXiv:2006.10934]
\bibitem[]{}Casella, P., Belloni, T., Homan, J., \& Stella, L. 2004, A\&A, 426, 587
\bibitem[]{}Casella, P., Belloni, T., \& Stella, L. 2005, ApJ, 629, 403
\bibitem[]{}Chaty, S., Munoz Arjonilla, A. J., \& Dubus, G. 2015, A\&A, 577 A101
\bibitem[]{}Corbel, S., Kaaret, P., Fender, R. P., Tzioumis, A. K., Tomsick, J. A., \& Orosz, J. A. 2005, ApJ, 632, 504
\bibitem[]{}de Ruiter, Iris, van den Eijnden, J., Ingram, A., \& Uttley, P. 2019, MNRAS, 485, 3834
\bibitem[]{}Done, C., \& Davis, S. W. 2008, ApJ, 683, 389
\bibitem[]{}Done, C., Gierlinski, M., \& Kubota, A. 2007, A\&ARv, 15, 1
\bibitem[]{}Doxsey, R., et al. 1977, IAUC, 3113, 2
\bibitem[]{}Emelyanov, A. N., Aleksandrovich, N. L., \& Sunyaev, R. A. 2000, Astron. Lett., 26, 297
\bibitem[]{}Espinasse, M., \& Fender, R. 2018, MNRAS, 473, 4122
\bibitem[]{}Ferreira, J., Petrucci, P. O., Henri, G., Sauge, L., \& Pelletier, G. 2006, A\&A, 447, 813
\bibitem[]{}Fender, R. P., Belloni, T., \& Gallo, E. 2004, MNRAS, 355, 1105
\bibitem[]{}Fender, R. P., Homan, J., \& Belloni, T. M. 2009, MNRAS, 396, 1370
\bibitem[]{}Fragile, P. C., Blaes, O. M., Anninos, P., \& Salmonson, J. D. 2007, ApJ, 668, 417
\bibitem[]{}Hawley, J. F., \& Krolik, J. H. 2019, ApJ, 878, 149
\bibitem[]{}Hjalmarsdotter, L., Axelsson, M., \& Done, C. 2016, MNRAS, 456, 4354
\bibitem[]{}Homan, J., Wijnands, R., van der Klis, M., et al. 2001, ApJS, 132, 377
\bibitem[]{}Homan, J., Miller, J. M., Wijnands, R., van der Klis, M., Belloni, T., Steeghs, D., \& Lewin, W. H. G. 2005, ApJ, 623, 383
\bibitem[]{}Homan, J., Bright, J., Motta, S. E., et al. 2020, ApJ, 891, 29
\bibitem[]{}Ingram, A., Done, C., \& Fragile, P. C. 2009, MNRAS, 397, L101
\bibitem[]{}Ingram, A., \& Done, C. 2012, MNRAS, 419, 2369
\bibitem[]{}Ingram, A., \& van der Klis, M. 2013, MNRAS, 434, 1476


\bibitem[Jahoda et al. 2006]{jahoda} Jahoda, K., Markwardt, C. B., Radeva, Y., et al. 2006, \apjs, 163, 401
\bibitem[]{}Jithesh, V., Maqbool, B., Misra, R., T, Athul. R., Mall, G., \& James, M. 2019, ApJ, 887, 101
\bibitem[]{}Kalamkar, M., Homan, J., Altamirano, D., van der Klis, M., Casella, P., \& Linares, M. 2011, ApJL, 731, L2
\bibitem[]{}Kaluzienski, L. J., \& Holt, S. S. 1977, IAUC, 3099, 3
\bibitem[]{}King, A., \& Nixon, C. 2018, ApJL, 857, L7
\bibitem[]{}Krimm, H. A., Barthelmy, S. D., Baumgather, W., et al. ATel, 2009, 2058, 1
\bibitem[]{}Kuulkers, E., Brandt, S., Budtz-Jorgensen, C., et al. ATel, 2008, 1739, 1
\bibitem[]{}Kuulkers, E., Kouveliotou, C., Belloni, T., et al. 2013, A\&A, 552, 32

\bibitem[]{}Kylafis, N. D., \&  Reig, P. 2018, A\&A, 614, L5
\bibitem[]{}Kylafis, N. D., Reig, P., \& Papadakis, I. 2020, A\&A, 640, L16
\bibitem[]{}Liska, M., Hesp, C., Tchekhovskoy, A., Ingram, A., van der Klis, M., \& Markoff, S. 2018, MNRAS, 474, L81
\bibitem[]{}Liska, M., Hesp, C., Tchekhovskoy, A., Ingram, A., van der Klis, M., \& Markoff, S. 2019, arXiv:1901.05970

\bibitem[]{}Liska, M., Hesp, C., Tchekhovskoy, A., Ingram, A., van der Klis, M., Markoff, S. B., \& Van Moer, M. 2020, accepted in MNRAS, staa009
\bibitem[]{} Lubow, S. H., Ogilvie, G. I. \& Pringle, J. E. 2002, MNRAS, 337, 706
\bibitem[]{}Makishima, K., Maejima, Y., Mitsuda, K., et al. 1986, ApJ, 308, 635
\bibitem[]{}Marcel, G., Cangemi, F., Rodriguez, J., Neilsen, J., Ferreira, J., Petrucci, P. O., Malzac, J., Barnier, S., \& Clavel, M. 2020, A\&A, 640, 18
\bibitem[]{}Markwardt, C. B., \& Swank, J. H. 2003, ATel, 133, 1
\bibitem[MR04]{MR04}McClintock, J. E., \& Remillard, R. A. 2004, in Compact Stellar X-ray Sources,
ed.W. H. G. Lewin \& M. van der Klis (Cambridge: Cambridge Univ. Press), 157
\bibitem[]{}McClintock, J. E., Remillard, R. A., Rupen, M. P., Torres, M. A. P., Steeghs, D., Levine, A. M., \& Orosz, J. A. 2009, ApJ, 698, 1398

\bibitem[]{}Merloni, A., Fabian, A. C., \& Ross, R. R. 2000, MNRAS, 313, 193
\bibitem[]{}Miller-Jones, J. C. A., Sivakoff, G. R., Altamirano, D., et al. 2012, MNRAS, 421, 468
\bibitem[]{}Miyamoto, S., Kimura, K., Kitamoto, S., Dotani, T., \& Ebisawa, K. 1991, ApJ, 383, 784
\bibitem[]{}Motta, S., Munoz-Darias, T., \& Belloni, T. 2010, MNRAS, 408, 1796
\bibitem[]{}Motta, S., Munoz-Darias, T., Casella, P., Belloni, T., \& Homan, J. 2011, MNRAS, 418, 2292
\bibitem[]{}Motta, S. E., Casella, P., Henze, M., Munoz-Darias, T., Sanna, A., Fender, R., \& Belloni, T. 2015, MNRAS, 447, 2059
\bibitem[]{}Motta, S. E., Casella, P.,  \& Fender, R. P. 2018, MNRAS, 478, 5159
\bibitem[]{}Nealon, R., Price, D., \& Nixon C. J. 2015, MNRAS, 448, 1526
\bibitem[]{}Nespoli, E., Belloni, T., Homan, J., et al. 2003, A\&A, 412, 235
\bibitem[]{}Nixon, C. J., King A. R., \& Price, D. J. 2012, MNRAS, 422, 2547
\bibitem[]{}Nixon, C., King, A., \& Price, D. 2013, MNRAS, 434, 1946
\bibitem[]{}Radhika, D., Nandi, A., Agrawal, V. K., \& Seetha, S. 2016, MNRAS, 460, 4403
\bibitem[]{}Reig, P., \& Kylafis, N. D. 2015, A\&A, 584, A109
\bibitem[]{}Reig, P., \& Kylafis, N. D. 2019, A\&A, 625, A90
\bibitem[]{}Remillard, R. A., Sobczak, G. J., Muno, M. P., \& McClintock, J. E. 2002, ApJ, 564, 962
\bibitem[MR04]{MR04}Remillard, R. A., \& McClintock, J. E. 2006, ARA\&A, 44, 49
\bibitem[]{}Remillard, R. A., McClintock, J. E., Orosz, J. A., \& Levine, A. M. 2006, ApJ, 637, 1002
 \bibitem[]{}Revnivtsev, M., Chernyakova, M., Capitanio, F., et al. 2003, ATel, 132,1
\bibitem[]{}Reynolds, M. T., \& Miller, J. M. 2013, ApJ, 769, 16
 \bibitem[]{}Reynolds, A. P., Parmar, A. N., Hakala, P. J., et al. 1999, A\&AS, 134, 287
 \bibitem[]{}Rupen, M. P., Mioduszewski, A. J., \& Dhawan, V. 2003, IAUC, 8105, 3
\bibitem[]{}Soleri, P., Belloni, T., \& Casella, P. 2008, MNRAS, 383, 1089

\bibitem[]{}Sriram, K., Agrawal, V. K., \& Rao, A. R. 2009, RAA, 9, 901
\bibitem[]{}Sriram, K., Rao, A. R., \& Choi, C. S. 2012, A\&A, 541, A6
\bibitem[]{}Sriram, K., Rao, A. R., \& Choi, C. S. 2013, ApJ, 775, 28
\bibitem[]{}Sriram, K., Rao, A. R., \& Choi, C. S. 2016, ApJ, 823, 67
\bibitem[]{}Steeghs, D., Miller, J. M., Kaplan, D., \& Rupen, M. 2003, ATel, 146, 1
\bibitem[]{}Steiner, J. F., McClintock, J. E., \& Reid, M. J. 2012, ApJL, 745, L7
\bibitem[]{}Steiner, J. F., Narayan, R., McClintock, J. E., \& Ebisawa, K. 2009, PASP, 121, 1279
\bibitem[]{}Stella, L., \& Vietri, M. 1998, ApJL, 492, L59
\bibitem[]{}Stevens, A. L., \& Uttley, P. 2016, MNRAS, 460, 2796
\bibitem[]{}Stevens, A. L., Uttley, P., Altamirano, D., et al. 2018, ApJL, 865, L15
\bibitem[]{}Tagger, M., \& Pellat, R. 1999, A\&A, 349, 1003
\bibitem[]{}Titarchuk, L. 1994, ApJ, 434, 570 
\bibitem[]{}Varniere, P., \& Tagger, M. 2002, A\&A, 394, 329
\bibitem[]{}Varniere, P., \& Vincent, F. H. 2016, A\&A, 591, 36
\bibitem[]{}van den Eijnden, J., Ingram, A., \& Uttley, P. 2016, MNRAS, 458, 3655
\bibitem[]{}van den Eijnden, J., Ingram, A., Uttley, P., Motta, S. E., Belloni, T. M., \& Gardenier, D. W. 2017, MNRAS, 464, 2643
\bibitem[]{}Wijnands, R., Homan, J., \& van der Klis, M. 1999, ApJL, 526, L33
\bibitem[]{}Xu, Y., et al. 2019, ApJ, 879, 93
\bibitem[]{}Yamaoka, K., Allured, R., Kaaret, P., et al., 2012, PASJ, 64, 32
\bibitem[]{} Zhou, J. N., Liu, Q. Z., Chen, Y. P., Li, J., Qu, J. L., Zhang, S., Gao, H. Q., \& Zhang, Z. 2013, MNRAS, 431, 2285


\end{thebibliography}
\end{document}